\documentclass{emulateapj}

\newcommand{\msun}{\rm M_\sun}
\newcommand{\hmsun}{h^{-1}\,{\rm M_\sun}}

\shorttitle{High-$z$ Galaxy Evolution: Effects of Galactic Winds}
\shortauthors{Sadoun et al.}

\begin{document}

\title{The Baryon Cycle at High Redshifts: Effects of Galactic Winds 
on Galaxy Evolution in Overdense and Average Regions}

\author{Raphael Sadoun\altaffilmark{1,2,6}}
\email{raphael.sadoun@utah.edu}
\author{Isaac Shlosman\altaffilmark{2,3}}
\author{Jun-Hwan Choi\altaffilmark{4,2}}
\author{Emilio Romano-D\'{\i}az\altaffilmark{5,2}}

\altaffiltext{1}{Department of Physics \& Astronomy, University of Utah, Salt Lake City, UT 84112-0830, USA}
\altaffiltext{2}{Department of Physics \& Astronomy, University of Kentucky, Lexington, KY 40506-0055, USA}
\altaffiltext{3}{Theoretical Astrophysics, Department of Earth \& Space Science, Osaka University, Osaka 560-0043, 
Japan}
\altaffiltext{4}{Department of Astronomy, University of Texas, Austin, TX 78712, USA}
\altaffiltext{5}{Argelander Institut fuer Astronomie, Auf dem Haegel 71, D-53121 Bonn, Germany}
\altaffiltext{6}{Institut d'Astrophysique de Paris (UMR 7095), 75014 Paris, France}

\begin{abstract}
We employ high-resolution cosmological zoom-in simulations focusing on a high-sigma peak and
an average cosmological field at $z\sim 6-12$, in order to investigate
the influence of environment and baryonic feedback on galaxy evolution in the reionization epoch.
Strong feedback, e.g., galactic winds, caused by elevated star formation rates (SFRs)
is expected to play an important role in this evolution.
We compare different outflow prescriptions: (i) constant wind velocity (CW), (ii) variable wind 
scaling with galaxy properties (VW), and (iii) no outflows (NW). The overdensity leads to
accelerated evolution of dark matter and baryonic structures, absent in the ``normal'' region,
and to shallow galaxy stellar mass functions at the low-mass end. Although CW shows little dependence on
both environments, the more physically motivated VW model does exhibit this effect. In addition, VW
can reproduce the observed specific SFR (sSFR) and the sSFR-stellar
mass relation, which CW and NW fail to satisfy simultaneously.
Winds also differ substantially in affecting the state of the 
intergalactic medium (IGM). The difference lies in volume-filling factor of 
hot, high-metallicity gas which is near unity for CW, while it remains confined 
in massive filaments for VW, and locked up in galaxies for NW.
Such gas is nearly absent in the normal region. Although all wind models suffer from deficiencies,
the VW model seems to be promising in correlating the outflow properties to those of host galaxies.
Further constraints on the state of the IGM at high-$z$ are needed to separate different wind models.
\end{abstract}


\section{Introduction}
\label{sec:intro}

With the rapidly increasing number of available multi-wavelength observations 
as well as the development of more sophisticated theoretical and numerical tools,  
our understanding of galaxy formation and evolution has progressed considerably in 
the recent years. While the majority of large galaxy surveys have focused 
on low and intermediate redshifts, substantial data are now available
to gain insight into the early stages of galaxy formation at $z>6$. 
In particular, observations of high-redshift quasars (QSOs) at $z\ga 6$ 
have hinted at the presence of rare, overdense regions hosting massive structures 
less than one billion years after the Big Bang \citep{Fan.etal:01}. 
Deep observations with the wide-area Suprime-Cam of the Subaru telescope have led to 
the discovery of $\sim 3\sigma$ overdensities of Lyman Break Galaxies
(LBGs) on scales of a few Mpc around the $z\sim 6.4$ QSO CFHQS J2329-0301 
\citep{Utsumi.etal:10}. Similar results have been obtained using the Large Binocular 
Telescope to study the environment of other high-$z$ QSOs
and also concluded that they reside in overdense regions \citep[e.g.,][]{Morselli.etal:14}. 
Recent theoretical work has shown that these regions do not necessarily evolve 
into the most massive clusters at $z=0$ and argued that galaxy formation might proceed 
differently there than in average density fields 
\citep{Trenti.Stiavelli:08,Overzier.etal:09,Romano-Diaz.etal:11a,Romano-Diaz.etal:11b}.

Star formation and AGN activity both appear to peak around $z\sim 2-3$ 
\citep[e.g.][]{Madau.Dickinson:14}, 
and galactic outflows, or galactic winds, 
are expected to exhibit a similar behavior, when averaged over a large enough volume. 
However, overdense regions, such as those hosting high-$z$ QSOs, have been predicted to evolve ahead of 
the average and underdense regions 
\citep[e.g.,][]{Barkana.Loeb:04,Romano-Diaz.etal:11a,Romano-Diaz.etal:11b,
Romano-Diaz.etal:14,Yajima.etal:15}. 
Although their evolution is concurrent with the reionization epoch at $6\la z\la 14$, they can 
be reionized ahead of the normal regions in the universe because of the elevated star formation rates
there. Moreover, these regions should collapse early, and, because the universe is also substantially 
denser at these redshifts, one expects mass accretion rates to be high.  
The question then is to what degree the associated galactic outflows from these objects 
affect their evolution as well as the evolution of their environment. 
In this paper, we aim to study the effects of these winds on galaxy evolution using 
high-resolution cosmological simulations of overdense regions and compare them with the evolution 
in average-density regions.

Massive outflows are projected to play an important role in the evolution of a large number 
of astrophysical objects --- from single and binary stars to supermassive black holes (SMBHs) 
and starburst galaxies. On galactic scales, winds are expected to affect the evolution of their hosts 
by regulating the star formation rate \citep[SFR, 
e.g.,][]{Scannapieco.etal:06,Oppenheimer.Dave:06,Schaye.etal:10}, 
ejecting gas and metals into the parent dark matter (DM) halos and the intergalactic medium 
\citep[IGM, e.g.,][]{Tremonti.etal:04,Dave.etal:08,Kirby.etal:08,Sijacki.etal:09}, and by determining the 
physical properties (disk mass and size, chemical evolution, galaxy luminosity function, and 
cuspiness of DM halos) of these objects \citep[e.g.,][]{Dekel.Silk:86,El-Zant.etal:01,El-Zant.etal:04,
Romano-Diaz.etal:08}. 
These outflows powered by supernovae (SN) can be supplemented by a feedback from the accretion 
processes onto the central supermassive black holes (SMBHs), the so-called AGN winds 
\citep[e.g.,][]{Blandford.Payne:82,Shlosman.etal:85,Emmering.etal:92,Konigl.Kartje:94,Murray.etal:95,
Arav.etal:97,Cecil.etal:01,Proga:03} for which there exists numerous observational 
evidence \citep[e.g.,][]{Emonts.etal:05,Croston.etal:08,Cano-Diaz.etal:12,Combes.etal:13,Cicone.etal:14}. 
A compelling example which has demonstrated the necessity 
for such strong feedback(s) can be found in the comparison study of disk evolution in a
cosmological setting, with and without feedback \citep[e.g.,][]{Robertson.etal:04,Schaye.etal:15}, 
where in the absence of a feedback the gas quickly violates the Toomre criterion and
fragments. Without feedback there is an overproduction of metals, especially in
small galaxies. Even more revealing is the overcooling problem which leads
directly to the angular momentum catastrophe --- the gas falls into the subhalos, and 
hitchhikes to the bottom of the potential well of the parent galaxy \citep[e.g.,][]{Maller.Dekel:02}, 
possibly contributing to the overgrowth of classical bulges.

The presence of galactic outflows is supported by numerous observations in both
low- and high-redshift galaxies \citep[e.g.,][]
{Osterbrock:60,Lynds.Sandage:63,Mathews.Baker:71,Heckman.etal:90,Hummel.etal:91,Heckman:94,
Kunth.etal:98}, including the $z\sim 3-4$ LBGs 
\citep[e.g.,][]{Pettini.etal:00,Adelberger.etal:03,Shapley.etal:03}, gravitationally-lensed 
galaxies at
$z\sim 4-5$ \citep[e.g.,][]{Franx.etal:97}, luminous and ultra-luminous IR (ULIRGs) galaxies 
\citep[e.g.,][]{Smail.etal:03}, etc. About 75-100\% of ULIRGs are associated with such winds --- 
their strength appears to correlate with the star formation rates (SFRs), but saturates for the most   
luminous ULIRGs \citep[e.g.,][]{Martin:05,Veilleux.etal:05}. Galactic winds are typically 
characterized by a biconical symmetry with respect to the underlying galactic disks 
\citep[e.g.,][]{Veilleux.etal:94,Shopbell.Bland-Hawthorn:98}. Hot tenuous winds show evidence for 
associated cold  $\sim 10^4$\,K  and neutral 
clouds  \citep[e.g.,][]{Heckman:94,Stewart.etal:00} ---  so they represent 
a multiphase interstellar medium (ISM).
An important link with theoretical predictions has been made by recent observations 
revealing that outflow velocities appear to correlate with the
galaxy stellar mass or its SFR \citep[e.g.,][]{Martin:05}.

A number of driving mechanisms have been proposed for galactic winds, 
such as radiation pressure in the UV lines and on dust, thermal pressure from the SN and 
OB stellar heating \citep[e.g][]
{Larson:74,Castor.etal:75,Dekel.Silk:86,MacLow.McCray:88,Ostriker.McKee:88,Shull.Saken:95,
Costa.etal:14,Vogelsberger.etal:14}, 
and pressure from cosmic rays \citep[e.g.,][]{Uhlig.etal:12}. The contribution of AGN is debatable at 
present. The driving of the outflows by the SN and stellar winds occurs when ejecta of individual 
sources form a bubble of hot gas, $\sim 10^{7-8}$\,K,
which expands due to the strong overpressure down the steepest pressure gradient and enters
the `blow-out' stage. Besides the mass, energy and momentum injected by such winds, this is
probably the main way for the highly-enriched material to be placed into the halo and further
out into the IGM  \citep[e.g.,][]{Cen.etal:05}. 

Galactic bulges may provide a testing ground for our understanding of various mechanisms that 
regulate the star formation and angular momentum
redistribution in a forming disk galaxy. An unexpectedly large fraction, $\sim 76\%$, of massive, 
$\ga 10^{10}\,\msun$ galactic disks can be fit with the Sersic index $\la 2$,
and $\sim 69\%$ have a bulge-to-the-total disk mass (B/T) ratio of $\la 0.2$, both in 
barred and unbarred galaxies \citep[e.g.,][]{Weinzirl.etal:09}. Such low Sersic indexes represent
disky rather then spheroidal stellar distribution \citep[e.g.,][]{Kormendy.Kennicutt:04}. On the 
other hand, bulges obtained
in numerical simulations are dominated by massive spheroids. This contradiction becomes even 
stronger in the recent study of the bulge population within a sphere of 11\,Mpc radius around the
Milky Way
using Spitzer 3.6\,$\mu$m and HST data \citep[e.g.,][]{Fisher.Drory:11}. The dominant galaxy type 
in the local universe has been found to possess pure disk properties, i.e., having a disky 
bulge or being bulgeless. These results reinforce the opinion that additional physical 
processes are needed to explain much less massive spheroidal components in disk galaxies.

Overall, strong arguments exist in favor of a process which lowers the efficiency of gas-to-stars 
conversion \citep[e.g.,][]{Fukugita.etal:98}. Such a process (or processes)
may resolve the discrepancy between the observed galaxy luminosity function (LF), both for high- 
and low-mass galaxies, and the computationally-obtained DM Halo Mass Function (HMF). 
Simulations of baryon evolution within the $\Lambda$ Cold DM ($\Lambda$CDM) framework appear to overproduce
low and high-mass galaxies.
Feedback from stellar evolution in low-mass galaxies and from AGN in high-mass ones is expected to 
quench the star formation, depleting the host galaxy from its interstellar medium (ISM), 
and resolving this discrepancy
\citep[e.g.,][]{Khochfar.Silk:06,Naab.etal:07,Somerville.etal:08,McCarthy.etal:12,
Hilz.etal:13}, see also reviews by \citet{Veilleux.etal:05} and \citet{Shlosman:13}.

In this work we focus at rare overdense regions (by construction) in the universe which host massive
$\sim 10^{12}\,\hmsun$ DM halos by $z\sim 6$, possibly the host halos of QSOs, and 
quantify the evolution of DM and baryons in these regions, 
in comparison with the `normal', average density regions. 
We compare numerical models of galactic winds by means of high-resolution 
cosmological simulations of galaxy formation and evolution at these high redshifts, 
and analyze efficiency of these winds in
carrying metals outside galaxies, into the halo and IGM environment. The effect of these winds
on individual galaxies is discussed as well.

This paper is structured as follows. Section 2 describes the numerical methods, initial
conditions used, and the details of the wind models. Resulting halo and galaxy properties 
are given in section 3, and the wind effects on the large-scale environment 
are analyzed in Section 4. Discussion and conclusions are given in the last section.

\section{Simulations}
\label{sec:simulations}

\begin{table*}
\caption{Summary of the different simulation runs and their properties}
\label{table:sim_properties}
\centering
\begin{tabular}{lcccccccccc}
\tableline
\tableline
Name     & Initial & Wind & $N_{\rm eff}$\tablenotemark{a} & $R_{\rm zoom}$\tablenotemark{b} 
         & $R_{\rm inner}$ \tablenotemark{c}
         & $m_{\rm DM}$\tablenotemark{d} & $m_{\rm gas}$\tablenotemark{d} & $m_\star$\tablenotemark{d}           
         & $\epsilon_{\rm grav}$\tablenotemark{e}  & $v_{\rm w}$\tablenotemark{f}  \\
         & conditions & model           &             & ($h^{-1}\,{\rm Mpc}$) & ($h^{-1}\,{\rm Mpc}$)
         & ($10^5\,\hmsun$) & ($10^5\,\hmsun$) & ($10^5\,\hmsun$) 
         & ($h^{-1}\,{\rm kpc}$) & (${\rm km\,s^{-1}}$) \\
\tableline
CW  & CR   & SH03 & $2\times 512^3$ & 3.5 & 2.7 & 37.3 & 8.88 & 4.44 & 0.14 & 484                \\ 
VW  & CR   & CN11 & $2\times 512^3$ & 3.5 & 2.7 & 37.3 & 8.88 & 4.44 & 0.14 & 1-1.5$v_{\rm esc}$   \\
NW  & CR   & $-$  & $2\times 512^3$ & 3.5 & 2.7 & 37.3 & 8.88 & 4.44 & 0.14 & 0                  \\
UCW & UCR & SH03 & $2\times 512^3$ & 7.0 & 4.0 & 37.3 & 8.88 & 4.44 & 0.14 & 484                \\
UVW & UCR & CN11 & $2\times 512^3$ & 7.0 & 4.0 & 37.3 & 8.88 & 4.44 & 0.14 & 1-1.5$v_{\rm esc}$   \\
\tableline
\end{tabular}
\tablenotemark{a}{effective resolution in the highest refinement (zoom-in) region,}
\tablenotemark{b}{comoving radius of zoom-in region,}
\tablenotemark{c}{comoving radius of central region used to identify halos and galaxies,}
\tablenotemark{d}{mass resolution in the DM, gas and stellar component,}
\tablenotemark{e}{comoving gravitational softening length,}
\tablenotemark{f}{outflow velocity used in the galactic wind model}
\end{table*}

\subsection{The code}
\label{sec:code}

We use a modified version of the tree-particle-mesh Smoothed Particle Hydrodynamics 
(SPH) code \textsc{GADGET-3}, originally described in \citet{Springel:05}, in its conservative entropy 
formulation \citep{Springel.Hernquist:02}. Our conventional code includes radiative 
cooling by H, He, and metals \citep{Choi.Nagamine:09}, 
a recipe for star formation and SN feedback, a phenomenological model for galactic winds, and 
a sub-resolution model for the multiphase ISM \citep[][hereafter SH03]{Springel.Hernquist:03}. 
In the multiphase ISM model, star forming SPH particles contain the cold phase 
that forms stars and the hot phase that results from SN heating. The cold 
phase contributes to the gas mass, and the hot phase contributes to gas pressure.
Metal enrichment is also taken into account according to the recipe from SH03
\citep[see also][]{Choi.Nagamine:09} in which the metallicity increase of
star-forming gas particles is related to the the fraction of gas in the cold phase, 
the fraction of stars that turn into SN and the metal yield per SN explosion.
Although metal diffusion is not explicitly implemented, metals can still be transported
outside galaxies by wind particles and enrich the halos and the IGM.  

Since we only focus on $z\ga 6$, before the full reionization, and do not implement 
on-the-fly radiative transfer of ionizing photons, the UV background is not included in our 
simulations. One expects this omission to have a larger effect on the overdense regions.
To quantify this would require an introduction of additional free parameter(s), as the precise level of the
UV background is unknown at present. We, therefore, refrain from doing so. For the same reason
we have neglected the AGN feedback \citep[e.g.,][]{Sijacki.etal:09} which has been predicted to affect 
the more massive galaxies.

For the star formation, we use the ``Pressure model'' \citep{Choi.Nagamine:10} which 
implements the relationship between the gas surface density and its volume density using
\citet{Schaye.DallaVecchia:08} prescription. The characteristic time-scale for the SF becomes 
$\tau_{\rm SF}\sim {\rm A}^{-1} (1\,\msun\,{\rm pc}^{-2})^n\,(\gamma P/G)^{(1-n)/2}$, 
where ${\rm A} = 2.5\pm 0.7\,\msun\,{\rm yr}^{-1}$, $n=1.4\pm 0.15$, $\gamma = 5/3$ \citep{Kennicutt:98}, 
and $P$ is the total effective gas thermal pressure (including the contribution from both cold
and hot phases). This model reduces the high-$z$ SFR relative 
to the standard recipe described in SH03. Star formation is triggered when the gas density is above the 
threshold value $n_{\rm crit}^{\rm SF} = 0.6\,{\rm cm}^{-3}$ \citep[e.g.,][]{Springel:05,Choi.Nagamine:10,
Romano-Diaz.etal:11b}. The value of $n_{\rm crit}^{\rm SF}$ is based on the translation of the threshold 
surface density in the Kennicutt-Schmidt law, SFR$\sim \Sigma_{\rm gas}^\alpha$,
where SFR is the disk surface density of star formation, $\Sigma_{\rm gas}$ is the surface
density of the neutral gas, and $\alpha\sim 1-2$, depending on the tracers used and on the
relevant linear scales. 

\subsection{Galactic wind models}
\label{sec:winds}

We have considered three different galactic outflow models: 
a constant velocity wind (CW) model based on the method of SH03 and 
a variable velocity wind (VW) model from \citet[][hereafter CN11]{Choi.Nagamine:11}, 
supplemented with a model without outflows (NW). 
All of our three wind models share the same thermal feedback as in SH03.

\paragraph{No-Wind model (NW)}
This model has no kinematic feedback, but maintains thermal feedback by the SN.

\paragraph{Constant Wind model (CW)}
The galactic wind has been triggered by modifying the behavior of some 
gas particles into the `wind' particles. Those were not subject to hydrodynamical forces 
and have experienced the initial kick from the SN. 
All CW particles had the same constant velocity, 
$v_{\rm w} = 484\,{\rm km\,s^{-1}}$, and the same mass-loading
factor, $\beta_{\rm w} \equiv\dot M_{\rm w}/\dot M_{\rm SF}$, where $\dot M_{\rm w}$ is the
mass loss in the wind and $\dot M_{\rm SF}$ is the SFR. The mass loading factor 
has been fixed to a value $\beta_{\rm w} = 2$ in agreement with the CW prescription 
used in CN11.

\paragraph{Variable Wind model (VW)}
The variable wind model (VW) from CN11 has introduced a more flexible subgrid physics 
compared to the simple recipe of SH03. 
In this paper, we adopt the 1.5ME wind model described in CN11. Briefly,
the model assumes that all gas particles in a given galaxy have the same chance to become
part of the wind. The probability that the gas particle becomes a wind particle is based on the values of 
$\beta_{\rm w}$, galaxy mass and galaxy SFR.
The main parameters are the wind load $\beta_{\rm w}$, defined above, and the wind
velocity $v_{\rm w}$ --- both have been constrained by observations which
express these two parameters in terms of the host galaxy stellar mass, $M_\star$, and the 
galaxy SFR.

The wind velocity is calculated as a fraction of the escape
speed from the host galaxy, $v_{\rm w} = \zeta v_{\rm esc}$, where $\zeta = 1.5$ for
momentum-driven and $\zeta = 1$ for energy-driven winds in the current setting.
The empirical relation between galaxy SFR and $v_{\rm esc}$ becomes,
\begin{equation}
{\rm SFR} = 1.0\, \left(\frac{v_{\rm esc}}{130\,{\rm km\,s^{-1}}}\right)^3
\left(\frac{1+z}{4}\right)^{-3/2}\, \msun\,{\rm yr^{-1}},
\label{eq:sfr}
\end{equation}
which is consistent with observations \citep[e.g.,][]{Martin:05,Weiner.etal:09}. 
The mass loading factor, $\beta_{\rm w}$, is assumed to represent the energy-driven wind 
($\beta_{\rm w} \propto v_{\rm esc}^{-1}$)
in the low-density case, $n < n_{\rm crit}^{\rm SF}$, and the momentum-driven wind 
($\beta_{\rm w} \propto v_{\rm esc}^{-2}$) for $n > n_{\rm crit}^{\rm SF}$.
Hence, for the low density gas, i.e., away from starforming regions (but inside the 
galaxy!), we apply the 
parameters of the energy-driven wind for the gas particles. 
This procedure requires that simulations compute $\beta_{\rm w}$ and $v_{\rm w}$ using
an on-the-fly group finder, which, in our runs, is a simplified version of the SUBFIND
algorithm \citep{Springel.etal:01}.
  
When an SPH particle is converted to a wind particle, it receives a $v_{\rm w}$ kick  
and decouples from hydrodynamic forces. The direction 
of the kicks are chosen to be preferentially perpendicular 
to the angular momentum of the particle such that, overall, the winds 
are launched above or below the galactic plane. The wind particles are turned
back to normal SPH particles when the ambient gas density
becomes lower than $0.1n_{\rm crit}^{\rm SF}$, or when they travel a distance
longer than 20 kpc/h, whichever comes first.

\subsection{Initial conditions}
\label{sec:ics}

The initial conditions have been generated using the Constrained Realization (CR) method 
\citep[e.g.,][]{Bertschinger:87,Hoffman.Ribak:91,Romano-Diaz.etal:07} and are those used by 
\citet{Romano-Diaz.etal:11b}, being downgraded from $2\times 1024^3$ to $2\times 512^3$. 
A CR of a Gaussian field is a random realization of such a field constructed to obey a set 
of linear constraints imposed on the field. The algorithm is exact, 
involves no iterations and is based on the property that the residual of
the field from its mean is statistically independent of the actual numerical value of the 
constraints (for more details see \citet{Romano-Diaz.etal:11a}). The main advantage 
of such method is 
to bypass the sampling problem of highly overdense regions which are rare and thus require to 
simulate a large volume ($\ga {\rm Gpc}^3$) of the universe.

The constraints were imposed on to a 
grid of $1024^3$ within a cubic box of size $20\,h^{-1}{\rm Mpc}$ to create a DM halo seed 
of $\sim 10^{12}\,\hmsun$ collapsing by $z\sim 6$, according to the top-hat model. We assume 
the $\Lambda$CDM cosmology with WMAP5 parameters \citep{Dunkley.etal:09}, 
$\Omega_{\rm m}=0.28$, 
$\Omega_\Lambda=0.72$, $\Omega_{\rm b}=0.045$, and $h=0.701$, where $h$ is the Hubble 
constant in units of 
$100\,{\rm km\,s^{-1}\,Mpc^{-1}}$. The variance $\sigma_8=0.817$ of the density field convolved 
with the 
top hat window of radius $8h^{-1}$\,Mpc$^{-1}$ was used to normalize the power spectrum.
The overdensity in the CR models corresponds to $\sim 5\sigma$, with
respect to the average density of the universe.

We also evolved the same parent random realization of the density field used to construct the CR 
runs but without any imposed constraints to represent an average region of the universe. 
In order to compare the effect of winds in different environments, we have 
run the unconstrained simulation using the CW and VW models. 
In the following, we refer to individual runs by the name of the wind model used in the 
simulation (see Table \ref{table:sim_properties}) : CW, VW and NW for constrained runs (CR runs), 
and UCW and UVW for their unconstrained counterparts (UCR runs).

The evolution has been followed from $z=199$ down to $z=6$. 
Our simulations were ran in comoving coordinates and with vacuum boundary conditions. 
We used the multimass approach and have downgraded the numerical resolution outside the central region 
using 3 different resolution levels in order to speed up the computational time. 
The highest refinement region has a radius of $3.5\,h^{-1}{\rm Mpc}$, having an effective 
resolution of $2\times 512^3$ in DM and SPH particles. The total mass within the computational
box is $6.19\times 10^{14}\,\hmsun$ and the mass of the high-resolution region is $1.45\times 
10^{13}\,\hmsun$. Within this region, we obtain the 
particle mass of $3.73\times 10^6\,\hmsun$ (DM), $8.88\times 10^5\,\hmsun$ 
(gas) and $4.44\times 10^5\,\hmsun$ (stars). The gravitational softening is 
$\epsilon_{\rm grav}= 140\,h^{-1}{\rm pc}$ (comoving), which is about $20\,h^{-1}{\rm pc}$ in 
physical units at $z=6$. 

In order to avoid contamination of heavy particles from the outer refinement levels, we have focused 
on the central inner spherical region within radius of $2.7\,h^{-1}{\rm Mpc}$ for our CR 
simulations, 
and with $4\,h^{-1}{\rm Mpc}$ for the unconstrained (UCR) simulations. The size of the inner region 
in the UCR 
runs has been chosen by matching the cumulative mass profile with the CR models. 
The properties of the different simulation runs are summarized in Table \ref{table:sim_properties}.

Finally, the CW model is switched on at the initial redshift, while the VW model is run as NW (i.e without
outflows) until $z=12$, at which point the winds are switched on.
This is done in order to speed up the simulation until the SF starts to rise and the
SN feedback becomes important.

\begin{figure*}
\includegraphics[width=\columnwidth]{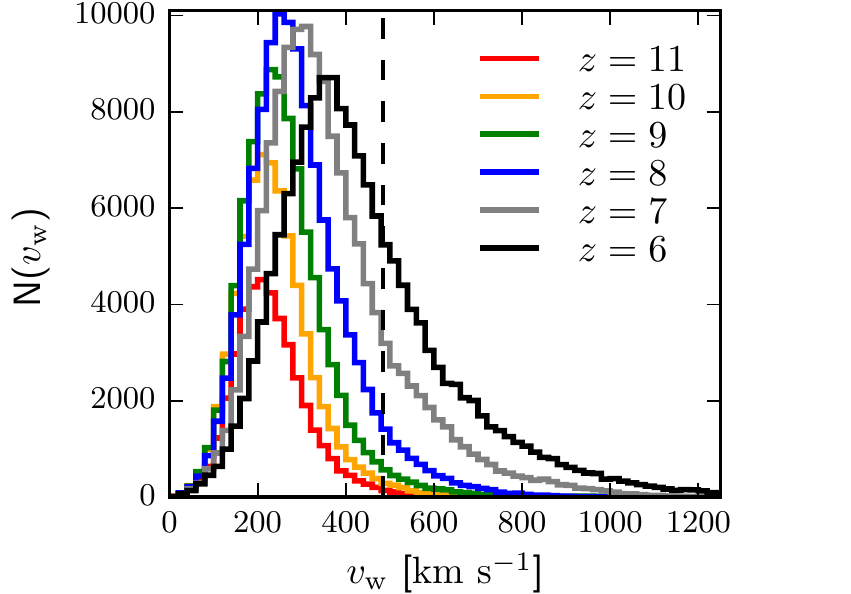}
\includegraphics[width=\columnwidth,height=0.7\columnwidth]{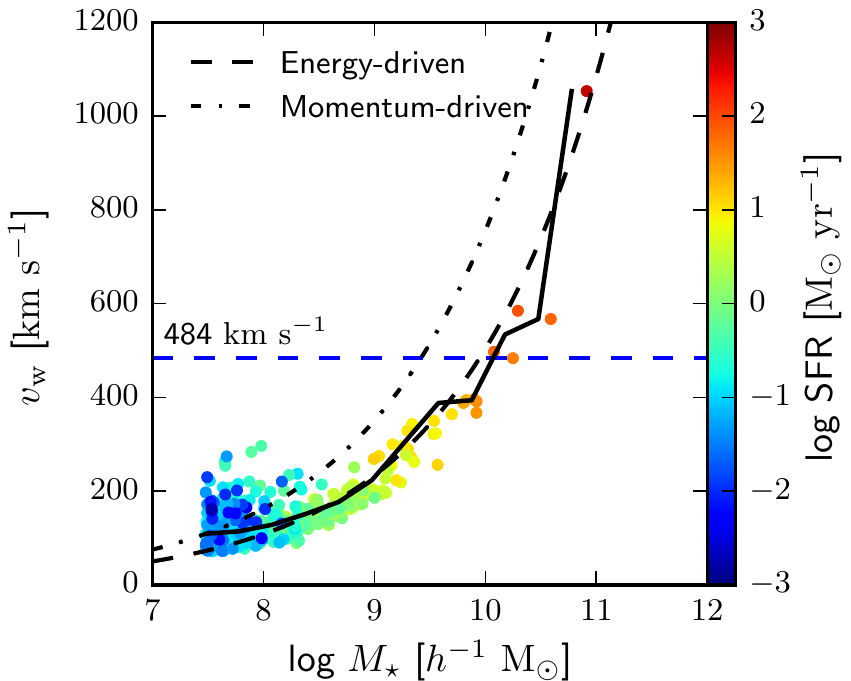} 
\caption{{\it Left:} Evolution of the distribution of wind particles as a function of their outflow 
velocities 
in the VW run. The vertical dash line corresponds to the fixed velocity of $484\,{\rm km\,s^{-1}}$ 
used in 
the CW model. {\it Right:} The median wind velocities as a function of galaxy stellar mass in the
VW model at $z=6$. The horizontal dashed line shows the velocity used in the CW model (as above). 
Filled circles represent individual galaxies in the VW run, color-coded by their median SFR. 
The solid line is showing the median relation between wind velocity and stellar mass 
for these galaxies, whereas the dashed and dotted line represents the expected trends
for energy-driven and momentum-driven winds respectively. \label{fig:vwind-vel}}
\end{figure*}

\subsection{Group finder}
\label{sec:hop}

\begin{table}
\centering
\caption{Properties of the halo and galaxy samples \label{table:halo_gal_samples}}
\begin{tabular}{lcccccc}
\hline
\hline
Name &  $M_{\rm h, min}$\tablenotemark{a} 
     &  $N_{\rm h, z=6}$\tablenotemark{b}
     &  $M_{\star, \rm min}$\tablenotemark{c} 
     &  $N_{\rm g, z=6}$\tablenotemark{d} \\
     &  ($\hmsun$) 
     &  
     &  ($\hmsun$)
     &  \\
         
\hline
CW  & $10^8$ &  6004 & $3 \times 10^7$ & 509 \\
VW  & $10^8$ &  5768 & $3 \times 10^7$ & 425 \\
NW  & $10^8$ &  5960 & $3 \times 10^7$ & 1969 \\
UCW & $10^8$ &  9135 & $3 \times 10^7$ & 245 \\
UVW & $10^8$ &  8470 & $3 \times 10^7$ & 92 \\ \hline
\end{tabular} \\
$^{\rm a}$Minimum halo total mass considered \\
$^{\rm b}$Number of halos at $z=6$ \\
$^{\rm c}$Minimum galaxy stellar mass considered \\
$^{\rm d}$Number of galaxies at $z=6$
\end{table}

We use the group finding algorithm HOP \citep{Eisenstein.Hut:98} to identify halos and galaxies.
Halos are isolated according to a purely particle density criteria --- the (local) \emph{total} 
particle densities, i.e., DM $+$ baryons, are calculated with an SPH kernel. A halo is defined as 
the region 
enclosed within a total iso-density contour of $\Delta_{\rm c}\rho_{\rm crit}(z)$, where 
$\Delta_{\rm c} = 80$ 
and $\rho_{\rm crit}(z)$ is the critical density at redshift $z$. Hence, no geometry 
of the particle distribution is assumed with this approach. This is a change compared to
\citet{Romano-Diaz.etal:11b} who used the standard pure DM definition for the halo, but is exactly as
in \citet{Romano-Diaz.etal:14}. The benefit of including the baryonic contribution to the density 
field when 
identifying halos, while not significantly modifying the resulting halo catalog, is to automatically 
identify 
the baryonic component inside each halo. In order to avoid resolution effects, we adopt a minimum mass 
threshold of $M_{\rm h} = 10^8\,\hmsun$.

Galaxies are identified with respect to the \emph{baryonic} density field with the outer boundary 
corresponding 
to an iso-density contour of $0.01 n_{\rm crit}^{\rm SF}$ which also differs 
from that of \citet{Romano-Diaz.etal:11b} but is as in \citet{Romano-Diaz.etal:14}. This definition 
ensures inclusion of regions which host star-forming gas ($n>n_{\rm crit}^{\rm SF}$), 
as well as lower density non-starforming gas ($n<n_{\rm crit}^{\rm SF}$), which is roughly bound to 
the galaxy. 

Following \citet{Romano-Diaz.etal:14}, we also exclude galaxies below a minimum \emph{stellar} 
mass of $3 \times 10^7\,\hmsun$, in agreement with recent observations \citep{Ryan.etal:14}, 
corresponding to $\sim 70$ stellar particles. 
In order to verify that our simulations are not affected by numerical resolution, we have
compared our CW run with the higher-resolution version presented in \citet{Romano-Diaz.etal:14}.
We found that both runs are in excellent agreement in terms of galaxy stellar mass functions,
star formation rates and metallicities in galaxies, indicating that our simulation runs
are numerically converged.

\section{Results: Effects of winds and environment 
on halo and galaxy population}
\label{sec:results}

\begin{figure*}
\epsscale{0.9}
\plotone{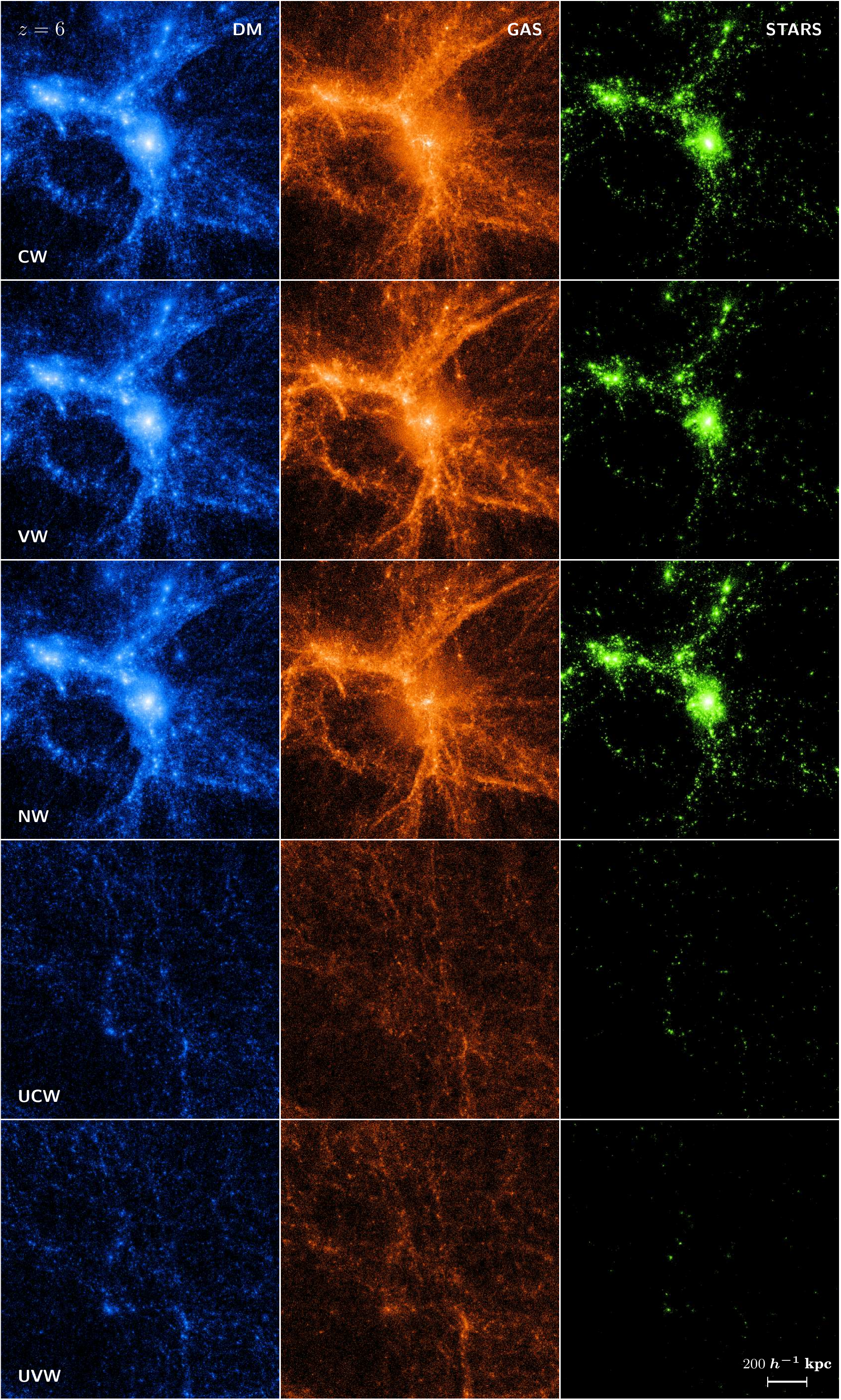} 
\caption{Snapshots of the mass distribution within the central $1.3\,h^{-1}{\rm Mpc}$ (comoving) 
at $z=6$ in the three different 
models considered here: Constrained Simulations (CR) with Constant Velocity (CW) wind (top row), 
Variable 
Velocity (VW) wind (second row), No wind (NW) (third row), and Unconstrained simulations with 
Constant Velocity (UCW) wind (fourth row) and Variable Velocity (UVW) wind (bottom row). 
Each column represents the view for a different component: Dark Matter (left column), gas 
(middle column) and 
stars (right column). In all panels, the same color scale represents column density in 
the corresponding component with an arbitrary normalization and is used only for visual 
representation of the mass distribution. \label{fig:frames}}
\end{figure*}

In this section, we compare the wind models in the overdense and average density 
environments and analyze their effects on the mass functions of halos and galaxies , as 
well as on the intrinsic properties of these objects, such
as gas and stellar fractions, star formation rates and on the mass-metallicity relation at
$z\ga 6$.

\subsection{Wind velocities}
\label{sec:windprops}

Section 2.2 provided the general description of the wind models implemented in this work.
We start by verifying the kinematic properties of these winds, and especially of the VW run.
In Figure~\ref{fig:vwind-vel}, we plot the resulting velocities of the wind 
particles in the VW and CW runs. The left panel exhibits the distribution of 
wind particles as a function of their outflow velocity, $v_{\rm w}$, 
at different redshifts. The vertical dashed line is the value 
used in the CW model, i.e., $484\,{\rm km ~s^{-1}}$.
Even though, overall, the total number of particles in the wind is steadily increasing with time, 
we observe three interesting trends in the VW model: the number of wind particles corresponding 
to velocity at the peak of the distribution is increasing with redshift until $z\sim 8$ and falls 
off thereafter; 
the average wind velocity increases with time and the tail of the distribution extends gradually towards 
higher velocities. Despite this latter increase, we note that, at all redshifts, 
the majority of wind particles in the VW run have lower velocity compared to the constant 
velocity adopted in the CW model.

The right panel of Figure~\ref{fig:vwind-vel} displays the relation between the median 
$v_{\rm w}$ and stellar mass in galaxies at $z=6$ in the VW run.
The value of $v_{\rm w}$ used in the CW model is shown as the horizontal dashed line. 
Individual galaxies are represented by filled circles with color scaling based on their SFR.
The solid black line is the median trend calculated from these objects. 
The black dot-dashed and dashed lines are the expected trends  
for the momentum-driven and energy-driven winds respectively. They have been
calculated from eq.~\ref{eq:sfr}, assuming a specific SFR (i.e., sSFR) of 2.5 $\rm Gyr^{-1}$, 
in accordance with results shown in Figure~\ref{fig:ssfr}. 

We find the scaling of wind velocities with stellar mass to closely 
follow the energy-driven case for nearly the entire mass range considered, although 
the scatter increases towards low masses. 
This is not entirely surprising since, in the VW model, winds 
coming from non star-forming regions ($n < n_{\rm crit}^{\rm SF}$) are assumed to 
be energy-driven. Given our definition of a galaxy, which includes gas well below 
the star formation threshold (section \ref{sec:hop}), these regions usually dominate the 
total gas content in a given galaxy. Since all gas particles have an equal probability 
to turn into wind, the majority of wind particles emitted from 
a galaxy come from those low-density regions and thus produce energy-driven winds, 
as seen on Figure \ref{fig:vwind-vel}.

We also find that, except for the most massive systems, nearly all galaxies in the VW model 
have lower median wind velocities than the CW case, in agreement with the velocity distributions 
presented in the left panel. The most massive objects appear to have stronger winds 
in the VW case than in the CW case. However, since they constitute the deepest potential wells, 
these objects are also the ones that are least affected by winds compared to the intermediate and 
low-mass galaxies. Thus, overall, we expect mechanical feedback 
from winds in the VW run to increase with time, peak at $z\sim 7-8$ and decline afterwards, 
as shown by the redshift evolution in the left panel. At the same time, this feedback is less 
efficient than in the CW run at all times. 
Given the direct coupling between wind and galaxy properties in the VW model, 
this non-monotonic evolution is accompanied with a similar behavior in galaxy gas content 
and SFR as we show below. In terms of stellar feedback, the VW model, therefore, represents 
an intermediate situation compared to the other two cases, NW and CW. \\

\begin{figure*}
\epsscale{0.9}
\plotone{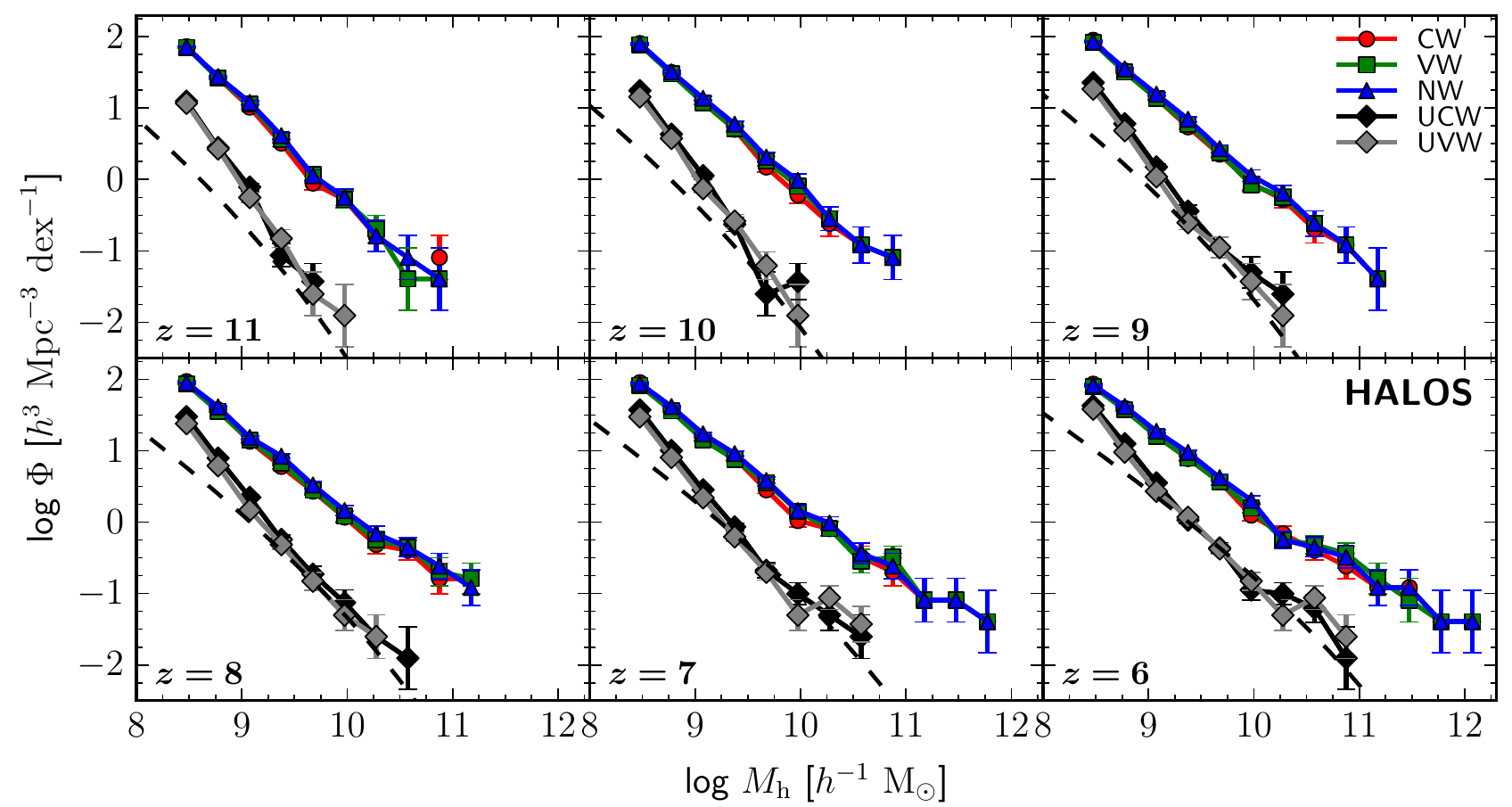}
\plotone{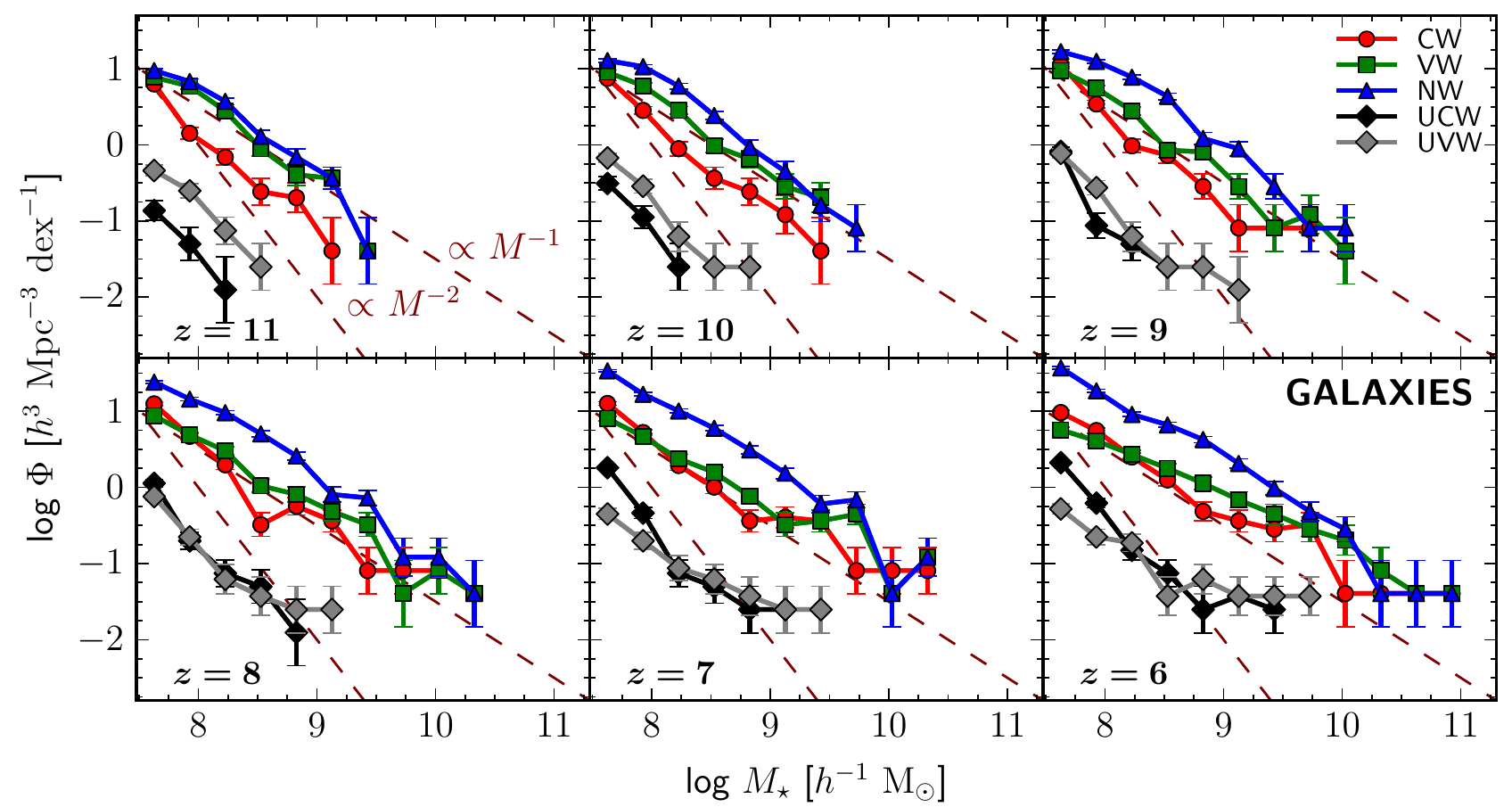}
\caption{{\it Top:} Evolution of the halo mass functions (HMFs) for the 
three CR wind runs considered here (CW, VW and NW) and for the 
UCR runs (UCW and UVW) at different redshifts. 
$M_{\rm h}$ is the total halo mass including DM$+$baryons. 
The dashed line corresponds to the theoretical Sheth-Tormen HMF. 
HMFs are insensitive to the outflow model used but show an overproduction 
of structures in the high-density region with respect to the average.
{\it Bottom:} Evolution of the galaxy mass function (GMF), as a function of their stellar mass, for  
CR and UCR runs. 
The dashed lines show the trend expected from a power-law relation 
scaling respectively as $M^{-1}$ and $M^{-2}$.
In both top and bottom panels, the mass functions are normalized by the binsize which is set 
to a constant 0.3 dex in $\rm log M$. Error bars correspond to the expected Poisson $1\sigma$ deviation 
in each bin. \label{fig:halo_gal_mfs}}
\end{figure*}

\subsection{Mass distribution and mass functions}
\label{sec:massfun}

Figure~\ref{fig:frames} shows the mass distribution of DM, gas and stars in all the simulation runs 
at $z=6$ on 
a scale of $1.3\,h^{-1}{\rm Mpc}$ (comoving), corresponding to a physical scale of 
$\sim 186\,h^{-1}$\,kpc. 
In the same environment (CR or UCR), 
no substantial differences are expected in the DM distribution on large scales since they only differ 
due to the physics of galactic outflows implemented.
On the other hand, we do observe that the distribution of baryons differs between the 
runs on progressively smaller spatial scales.
 
For a more quantitative analysis, we start by examining the differential HMFs, 
$\Phi = dn/d{\rm log}\,M_{\rm h}$, i.e., the number of halos 
per unit volume per unit logarithmic total halo mass interval, as shown in 
Figure~\ref{fig:halo_gal_mfs} (upper frame). 
As expected, the HMFs of the different wind models in a given environment (CR or UCR) are essentially 
identical 
over the entire mass range of $10^8\,\hmsun - 10^{12}\,\hmsun$. Slight differences are observed 
between wind 
models because our definition of a halo is based on the {\it total mass which includes baryons} 
(section\,\ref{sec:hop}).

On the other hand, the effect of the overdensity is readily visible from this figure.
First, the HMFs for the three CR models are shifted up with respect to the UCR HMFs
because of the overdense region sampled by these models. Second, the slopes of the
three models are also somewhat shallower than that of the theoretical Sheth-Tormen slope 
\citep{Sheth.Tormen:99}, as expected
around density peaks \citep[e.g.,][]{Romano-Diaz.etal:11a}. Third, they extend toward higher masses,
compared to the UCR halos. The HMFs in the UCR runs appear to be in reasonable 
agreement with the \citeauthor{Sheth.Tormen:99} HMF, except at the low-mass end, 
where we see some deviation from the theoretical prediction.

We have analyzed the possible causes of this excess of low-mass halos in the UCR simulations
over the Sheth-Tormen HMF. First, it could follow from the spurious detection of gas-dominated blobs
which might be identified as low-mass halos by HOP. We, therefore, have computed the baryon fraction 
$f_{\rm b} = M_{\rm baryons}/M_{\rm DM}$ for each halo in the UCR runs at z=6, but found 
that $f_{\rm b}$ is always below 40\%. Second, the difference is still present even
when we use only the DM component to find and define our halos. Third, as a complementary test, we 
have also analyzed the HMFs
in a {\it DM-only} version of our UCW model (run in a larger volume) presented in 
\citet{Romano-Diaz.etal:11a} and found a good agreement with the theoretical HMFs from 
\citet{Jenkins.etal:01}
over the entire mass range considered here. For this reason, we are confident
that the departure between our HMFs and the theoretical prediction from \citeauthor{Sheth.Tormen:99}
is due to the inclusion of baryonic physics, which is expected to have the most effect on 
low-mass halos (in preparation).
  
The bottom panel of Figure~\ref{fig:halo_gal_mfs} 
shows the galaxy mass functions (GMFs) for all models as a function of galaxy
stellar mass, $M_\star$, at different redshifts. 
Here we observe a clear impact of the outflow model on the shape and evolution 
of the GMF, reflecting the effect of winds on the galaxy growth process. 
At every redshift, the normalization (amplitude) of the GMF in the NW model is the 
highest overall among the CR runs, which means that galaxies of a given stellar mass are more 
abundant in the NW model compared to other two CR runs (CW and VW). However, since the CR 
runs have the same underlying HMFs, an alternative way to characterize the differences between GMFs 
is to 
compare their relative shift at a given number density $\Phi$ (i.e at a given halo mass). 
In this case, we find that the GMF in the NW model is shifted towards higher stellar masses, 
meaning that, \emph{in a given halo}, galaxies have produced more stars in the NW case because 
of the absence of feedback from winds with respect to the models including outflows (CW and VW). 

In comparison, the effect of winds is prominent in the evolution of the VW and CW GMFs 
at different redshifts. At $z=11$, VW and NW GMFs closely follow each other as 
expected since these two models are identical until 
winds are turned on at $z\sim 12$ in the VW run (section \ref{sec:winds}). 
With decreasing redshift, the outflows become stronger which cause both VW and 
CW GMFs to become shallower, especially at the low-mass end. At $z=6$, the VW GMF is more closely 
related 
to the CW one, but their slopes differ, which results in a smaller number of low-mass ($\sim 10^8\, 
\hmsun$) galaxies and a larger population of objects at intermediate masses in the VW run. 
The differences mentioned above are no longer visible at the massive end 
($M_\star \ga 10^{10}\, \hmsun$) of the GMFs where all CR runs agree reasonably well. This means 
that winds are inefficient in removing baryons from the deepest potential wells and thus 
affect mostly the low-mass end of the GMFs, in agreement with previous models and observations
\citep[e.g.,][]{Benson.etal:03,DeYoung.Heckman:94,MacLow.Ferrara:99,Scannapieco.etal:01,
Choi.Nagamine:11}.

The evolution of the GMFs in the average density region (UCW and UVW models) 
appears to resemble their wind model counterparts (CW and VW) in the CR runs, but shifted towards 
lower stellar masses. 
In particular, we observe a flattening with time of the low-mass end of the UVW GMF compared to the 
UCW case. 
However, the main difference between the two environments lies in the low-mass end of the GMFs 
which appear to be steeper for UCR than for CR runs. The difference is especially pronounced when 
comparing 
CW and UCW runs which have low-mass end slopes of $\sim -2$ and $\sim -1$ respectively. 
Because UCR and CR runs only differ by the presence of the imposed overdensity, it is clear that these 
trends reflect the difference in the underlying HMFs, as seen in the top panels
of Figure~\ref{fig:halo_gal_mfs}. This is an important result concerning 
galaxy evolution in overdense environments at high-redshifts which should have implications for the 
expected contribution of low-mass galaxies to reionization at $z \ga 6$, 
since measurements of the faint-end slope of the UV LF at high-$z$  indicate a steep trend 
with a slope of $\sim -2$ \citep{Dressler.etal:15,Song.etal:15} in agreement with the trend we find 
in the average density regions.

\subsection{Gas fractions: effect of winds}
\label{sec:fractions}

\begin{figure}
\plotone{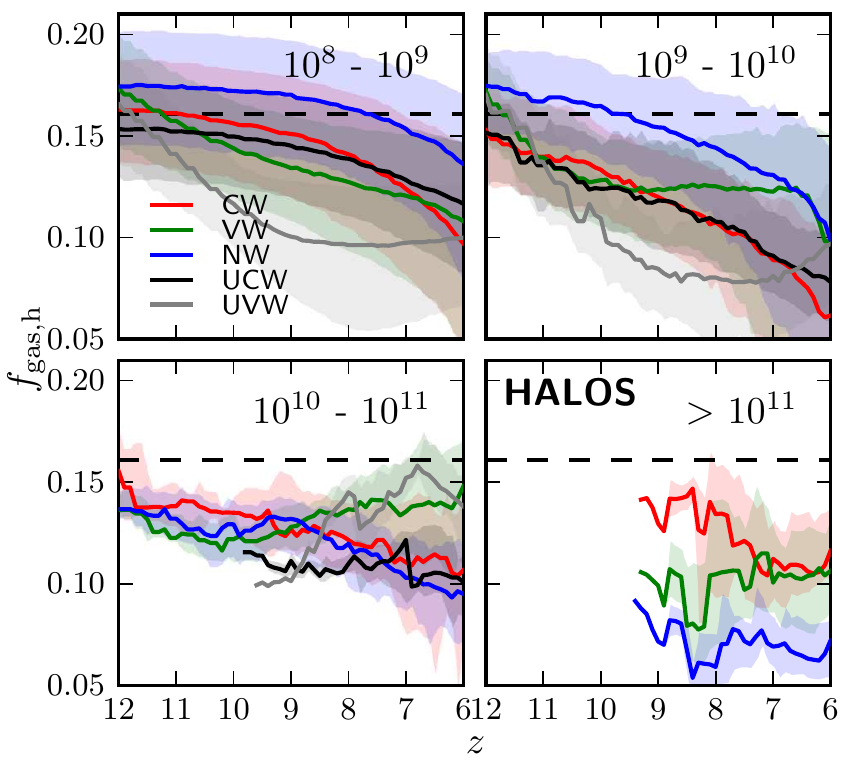}
\plotone{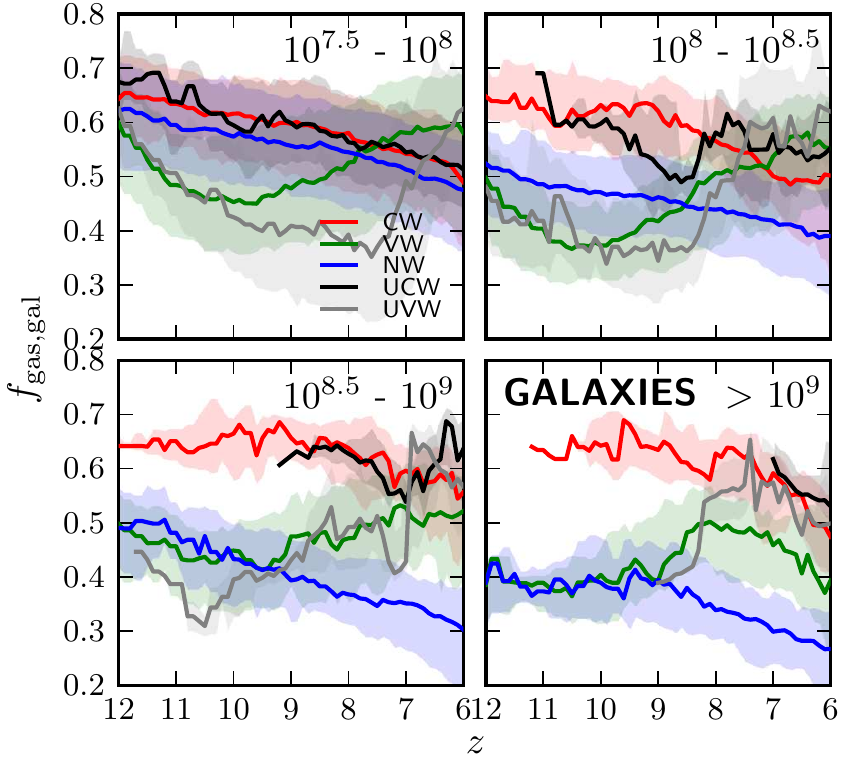}
\caption{Evolution of gas fraction in halos (\emph{top}) 
and galaxies (\emph{bottom}) as a function of redshift 
in different mass bins. Objects are assigned to one of the 
bins based on their \emph{current} total mass (halos) or stellar 
mass (galaxies) at each redshift. Bin limits are indicated 
in each panel in units of $\hmsun$. Solid lines 
are the median trends in each bin. 
Shaded regions show the scatter around the median between the 
20 and 80 percentiles of the population.
In the top figure, the dashed horizontal line indicates the
universal baryon fraction $\Omega_{\rm b}/\Omega_{\rm m}$.
\label{fig:fgas_halos}}
\end{figure}

Next, we analyze the impact of the different galactic wind prescriptions 
on the gas content of halos and galaxies.
Figure~\ref{fig:fgas_halos} (top) displays the halo gas fraction, $f_{\rm gas,h} = M_{\rm
gas,h}/M_{\rm h}$ as a function of redshift in four mass bins, from 
$M_{\rm h} = 10^8$ to $10^{12}\,\hmsun$ with a bin size of 1~dex. 
We keep these mass bins constant in time, which means that halos are 
assigned to one of the bin based on their total mass at a 
given redshift. In each panel, solid lines represent the median trend of each model, 
and shaded regions enclose the 20 and 80 percentiles of the population in each bin.

For all models, we observe that the evolution of the gas content in halos shows a clear 
dependence with halo mass. Low-mass halos start with a gas fraction close to the 
universal baryon fraction, $f_{\rm bar,0} = \Omega_{\rm b}/\Omega_{\rm m}\sim 16\%$ (dashed line), 
and exhibit a steeper decline with $z$ than the more massive ones. As halo mass increases, the curve 
flattens, with the highest-mass bin, $\ga 10^{11}\,\hmsun$, showing little 
variations in gas fraction, from $z=9$ to $z=6$, although trends are more noisy in this bin 
because of the low number of objects at each $z$. At later times, $f_{\rm gas,h}$ is 
in the range of $\sim 5-15\%$, below the universal value $f_{\rm bar,0}$. 
Note that the highest mass bin does not contain any UCR halos since such massive objects 
have not formed yet in the average density region.

The CR runs (CW, VW and NW) show clear differences among themselves, 
reflecting the impact of the outflow model on the gas content of halos. 
On average, at each $z$, gas fractions are highest in the NW model for lower mass halos, 
$\la 10^{10}\,\hmsun$,  and become 
lower than in the CW and VW models for the most massive objects, $\ga 10^{11}\,\hmsun$. 
Although it seems contradictory at first that NW halos have less gas, we find this trend 
to be accompanied by a similar trend in SFRs. 
The SF is more efficient and starts earlier in the NW model because of the absence of a substantial 
feedback from galactic outflows.
The SFRs are also mass-dependent with very little SF at the low-mass end in all models. 
As a consequence, since halos in the NW run are able to better retain their gas 
compared to the CW and VW, more of it gets consumed to form stars in higher mass objects 
which explains the above trend. Such a high efficiency in converting the gas into stars is
one of the reasons that the NW model can be ruled out.

Figure~\ref{fig:fgas_halos} (bottom)
displays the evolution of the gas fraction, $f_{\rm gas,gal} = M_{\rm gas}/(M_{\rm gas}+M_\star)$,
in galaxies in four mass bins. The lines and shaded areas 
have the same meaning as in the case for halos (top) but now the galaxies 
are binned in terms of their stellar mass $M_\star$ at a given $z$. 

The comparison between the CR models shows a diverse population of objects whose gas fraction evolution
is much more complex than in halos. The NW model shows a monotonic decline of $f_{\rm gas,gal}$
with time, except in the highest mass bin for $z\ga 9$. By $z\sim 6$, the gas fraction
is decreasing with increasing mass bin. At the same time, $f_{\rm gas,gal}$ is independent of mass
in the CW model. For the VW model, the final $f_{\rm gas,gal}$ is also lower for higher galaxy masses. 
The overall evolution of CW and VW models, however, differs substantially. For almost the entire 
mass range,
the gas fraction stays constant for $z\ga 8-9$, and exhibit a mild decline thereafter.
On the other hand, $f_{\rm gas,gal}$ in the VW model shows a sharp decline for $z\ga 9$, followed
by a sharp increase, and saturation. In the highest mass bin, another sharp decline can be clearly
observed.

Overall, by $z\sim 6$, the NW galaxies appear to be most gas-poor, followed by VW galaxies and CW ones.
Obviously, this trend correlates with the efficiency of feedback in the form of galactic outflows.
The weakest feedback is that of the NW model. The CW feedback is strong but fixed in time, while 
that of the VW is increasing with time, but not monotonically. 
So averaged over the whole galaxy population, the effect of the VW model 
falls roughly between that of the CW and NW models as already mentioned in section \ref{sec:windprops}.
 
In comparison with \citet{Choi.Nagamine:11} galaxies, the spread between various models is smaller 
at $z\sim 5$, and $f_{\rm gas,gal}\sim M_{\rm *,gal}^{-1}$, with 
$f_{\rm gas,gal}\sim 0.9$ at ${\rm log}\,M_{\rm *,gal}\sim 7$. 
Our CR models show smaller gas fractions for all galaxies already at $z\sim 6$, which is the 
result of elevated SFR caused by the higher accretion rates and gas supply in the overdense region. 
This trend persists despite the overdense region, especially at the high mass end.
This is also true in comparison with \citet[]{Thompson.etal:14} models. The explanation for the gas 
fraction evolution that we observed in the CR models lies in the interplay between the SF and mass
accretion histories onto galaxies \citep[e.g.,][]{Romano-Diaz.etal:14}. As we discuss later on, 
the kinetic luminosity of these winds has a substantial effect on the temperature of the IGM and
the halo gas --- an effect which lowers accretion rates profoundly.

\subsection{Gas fractions: effect of environment}
\label{sec:fractions_env}

\begin{figure}
\epsscale{1.1}
\plotone{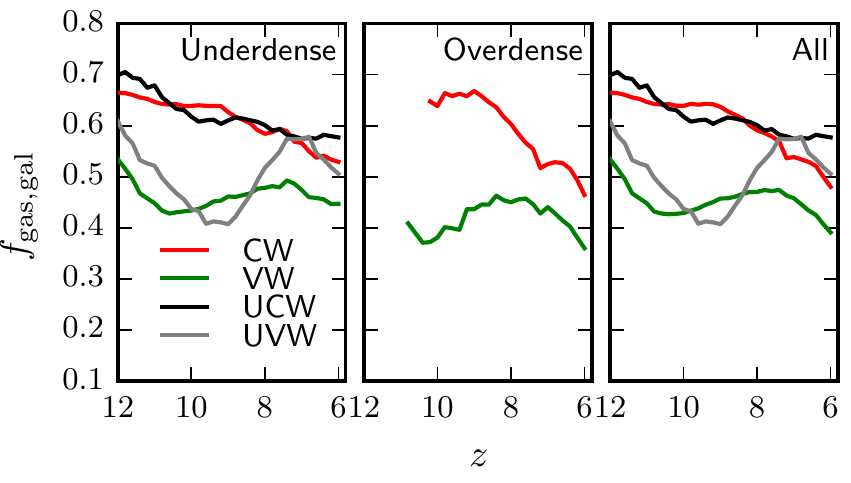}
\plotone{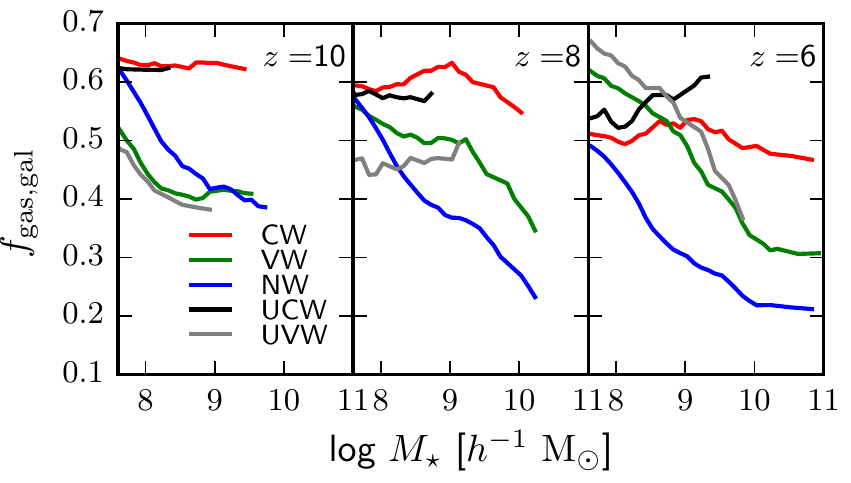}
\caption{\emph{Top}: Evolution of gas fraction in galaxies 
in the CR and UCR runs 
as a function of redshift, similar to Fig. \ref{fig:fgas_halos}, but 
objects are now binned by density 
contrast $\delta = \rho/\bar{\rho}_{\rm b}-1$ 
where $\bar{\rho}_{\rm b}$ is the universal mean baryonic density. 
The bins are defined as: $\delta < 0$ for underdense (left panel), 
$\delta > 0$ for overdense (middle panel) and all galaxies
(right panel).
\emph{Bottom}: Mass-weighted average gas fractions in galaxies as function of stellar mass 
at $z=10$ (left panel), $z=8$ (middle panel) and $z=6$ (left panel).
\label{fig:fgas_env}}
\end{figure}

As evident from Figure\,\ref{fig:fgas_halos}, the evolution of gas content 
in CW and UCW halos and galaxies is nearly identical in all mass bins, except in the
mass bin $10^{10}-10^{11}\,\hmsun$, where the most massive objects are found in the UCR runs.
Since both models use the same outflow prescription, this
strongly suggests that, for a given wind model, 
the overdensity has little effect on the amount 
of gas inside these halos and galaxies. 

When comparing VW and UVW halos, we find that their gas fractions tend to follow
similar trends in each mass bin, but with larger relative differences than between CW and UCW
runs.
This indicates that the VW model couples galaxy evolution with the large-scale environment more
strongly than the CW case.

However, addressing
the effect of the environment on gas fraction is difficult from a direct 
comparison between CR and UCR evolution from Fig. \ref{fig:fgas_halos} 
because objects in a same mass bin can reside in 
different environments. For example, low-mass galaxies can be found  
in underdense regions in CR and UCR models, but also as satellites 
around the massive halos in the overdense region. 
This effectively causes mixing among objects of similar masses and 
can thus wash out the differences induced by the presence of the overdensity.

To investigate further the effects of the environment on the gas content in 
galaxies, it is thus easier to directly divide objects depending 
on their large-scale environment. Figure~\ref{fig:fgas_env} (top) shows 
the redshift evolution of gas fractions in galaxies in different density bins for 
the CR and UCR models that include outflows.
The bins are defined based on the smoothed relative density contrast 
$\delta = \rho/\bar{\rho}_{\rm b}-1$ where $\bar{\rho}_{\rm b} = \Omega_b\rho_{\rm crit}$ 
is the universal mean baryonic 
density. The smoothed galaxy density field $\rho$ has been obtained by convolving the 
local galaxy densities calculated using an SPH method with a top-hat filter of 
$250\,h^{-1}\,{\rm kpc}$ scale. Note that we use the universal mean baryonic density 
$\bar{\rho}_{\rm b}$ 
in our definition of $\delta$ instead of the mean baryonic density inside the simulation box 
in order to have a common normalization factor for both CR 
and UCR models. Galaxies are divided between underdense ($\delta<0$, top left panel) 
and overdense ($\delta>0$, top middle panel) bins and we also show 
the evolution for the entire galaxy population for comparison (top right panel). 
Based on our galaxy density definition, we found no galaxies with $\delta>0$ 
in the UCR models. The comparison between CR and UCR runs 
is therefore only possible for galaxies residing in underdense regions. 

The top left panel of Figure~\ref{fig:fgas_env} shows that the evolution of gas fraction 
in these galaxies residing in underdense regions
follows a close trend in both CW and UCW models with a modest decline 
from $\sim 65$-$70$\% at $z=12$ to $\sim 55$\% at $z=6$. 
Differences between the two models are not significant given the noise introduced 
by the small number of galaxies formed in the UCW model. 
In comparison, the gas fraction in galaxies residing in overdense regions in the CW model 
exhibits a much steeper decline from $z=10$ to $z=6$ (top middle panel) due to the accelerated 
evolution in these regions. The UVW galaxies show a more complicated behavior,
but decline as well after $z\sim 8$, in tandem with CW.
The evolution for the entire galaxy population (top right panel) shows 
larger differences between CW and UCW 
due to the mixing of both underdense and overdense environments in the CW case 
but are still showing a similar decreasing 
trend with redshift and comparable gas fraction levels at all $z$. 
The VW and UVW models in the underdense bin exhibit 
much less correlation than CW and UCW among themselves, as already noted in the previous section.    

\begin{figure*}
\epsscale{0.95}
\plotone{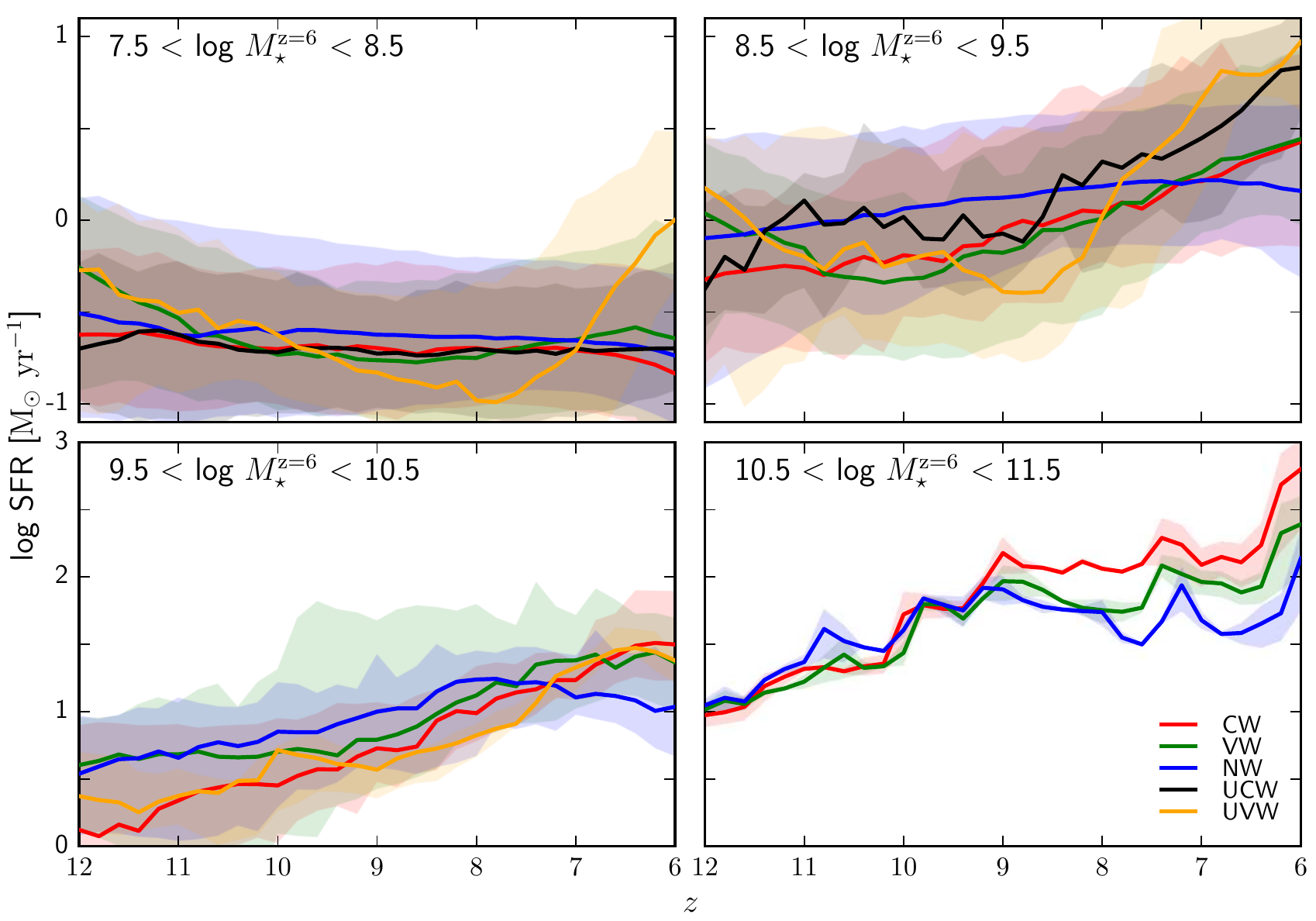}
\caption{Star formation histories for galaxies in different stellar mass bins. 
Galaxies are selected based on their stellar mass $M_\star^{z=6}$ at $z=6$ (which is 
different than selecting galaxies based on their mass at every redshift) and their 
ancestors are identified at earlier redshifts to be the most massive object among all the progenitors. 
The bin limits are indicated in each panel in units of $\hmsun$.
Solid lines show the median SFR evolution of all the ancestors at each $z$.
The shaded regions are bounded by the 20 and 80 percentiles.\label{fig:sfh}}
\end{figure*}

The similar gas fraction evolution for CW and UCW galaxies residing in the same environment
further confirms that the gas content is mainly affected by 
the different wind models and not by the presence of the overdensity. The gas fraction in
VW and UVW models differ more among themselves, which can be explained by weaker winds they produce.
To more fully understand 
the reason for this effect, we need to investigate how the galaxy 
gas fractions depend on the outflow prescription 
at a {\it given} redshift. Figure~\ref{fig:fgas_env} (bottom) shows the mass-weighted
average galaxy gas fractions as a function of galaxy stellar mass in all the simulation runs 
at $z=10$, 8, and 6. 
Since we are only interested in the average trends, the curves 
have been smoothed with a boxcar filter of size 0.5\,dex in stellar mass 
to reduce Poisson noise caused by the small number of objects in each mass bin. 

We find significant differences among the various wind models at all $z$ but 
the main discrepancy can be observed between CW and the other two runs, VW and NW. 
In particular, there is a clear trend of decreasing gas fraction with increasing 
stellar mass in both VW and NW, which is not seen in CW. The latter shows nearly 
constant gas fractions across the entire mass range $M_\star = 10^{7.5}$--$10^{11}\,\hmsun$. 
The UCW model also exhibits nearly mass-independent gas fractions at all $z$, which 
follows the CW trend as expected. This is in agreement with results shown 
in the top panels of Figure~\ref{fig:fgas_env} and Figure~\ref{fig:fgas_halos}, 
although there are clear signs of divergence between the two models for 
$z\le 8$.
These results thus suggest that the likely explanation for the apparent similarities in 
galaxy gas content between CW and UCW can be attributed, for the most part, 
to the constant and strong outflow prescription used in both runs. Since winds scale 
independently of galaxy properties (mass and SFR) in this outflow model, the gas content inside 
galaxies varies weakly with stellar mass. As a consequence, the CW and UCW 
runs produce galaxies with nearly constant gas fractions, which explains the similar behavior 
observed between the two models despite their different environments.

\subsection{Star formation histories}
\label{sec:sfhistory}

To construct the star formation history (SFH) for a given galaxy selected at $z=6$, we identify the 
progenitors at earlier redshifts by matching the particle unique identification 
number (IDs). The ancestor galaxy is taken to be the progenitor containing the largest 
number of particles. 
We repeat this process iteratively to obtain the list of ancestors of the selected galaxy, 
from $z=6$ to $z=12$. 
Figure~\ref{fig:sfh} shows the redshift evolution of galaxy SFRs in different mass bins. 
Galaxies are assigned to one of the bins based on their stellar mass $M_\star^{z=6}$ at $z=6$, 
and their 
SFH is then followed back to $z=12$ using the above method. 
We stress that this mass binning is different from the one in which galaxies are selected by mass 
at each redshift, as used in sections~\ref{sec:fractions} and \ref{sec:ssfr}. For this reason, attempting 
to correlate SFR and gas fraction evolution cannot be performed using a straightforward 
comparison between figures~\ref{fig:fgas_halos} and \ref{fig:sfh}, as they represent different galaxy 
populations at each redshift.

We find a strong dependence of SFHs on stellar mass in all models.
Overall, massive galaxies exhibit a strongly rising SFR, while the lowest mass galaxies
maintain flat SFRs on average. These trends are well correlated with 
the mass accretion rates onto the galaxies, analyzed in \citet{Romano-Diaz.etal:14}. 
More specifically, in the lowest mass range, $7.5<\log\,M_\star^{z=6}<8.5$, 
SFRs are essentially constant with $z$ for all models, and lie in the range of $\sim 0.1 - 
1\,\msun\,{\rm yr^{-1}}$. 
The intermediate mass bins, $8.5<\log\,M_\star^{z=6}<9.5$ and $9.5<\log\,M_\star^{z=6}<10.5$, 
exhibits an increase in SFR by a factor of $\sim 7-8$ between $z=12$ and $z=6$. 
Finally, the most massive galaxies display the strongest increase, roughly a factor of 30 over this 
redshift interval. Note, however, that small SF levels seen for low-mass galaxies  likely result 
from poorly resolved outflows, with only few particles for these objects. For this reason, 
we mostly focus on SFRs in intermediate and high-mass galaxies in the following.

\begin{figure*}
\plotone{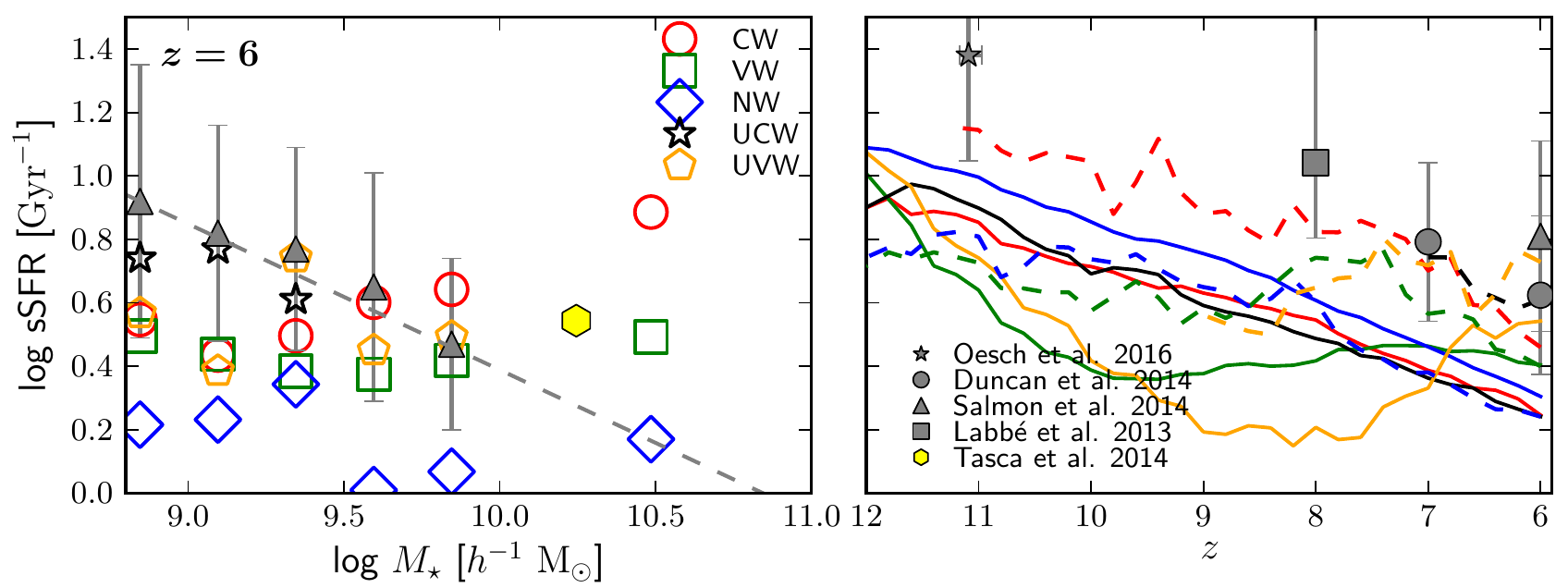}
\caption{Median specific star formation rate (sSFR) as a function of stellar mass (left) and redshift 
(right). Observational data points are taken from \citet{Labbe.etal:13} at $z=8$, \citet{Salmon.etal:15} 
at $z=6$ and \citet{Duncan.etal:14} at $z=6-7$. The
yellow hexagon (left frame) is an observational constraint at $z\sim 5-5.5$ from the VUDS survey 
(\citet{Tasca.etal:15}, see their Fig.\,3).
The sSFR-stellar mass relation in the left panel is shown for galaxies at $z=6$. The dashed grey 
line is derived from the fitted SFR-$M_\star$ relation at $z=6$ from \citet{Salmon.etal:15}.  
In the right panel, the solid lines correspond to the median sSFR of all the galaxies at each redshift. 
The dashed lines are the median sSFR for galaxies with stellar masses in the range 
$10^{9}-10^{9.5}\,\hmsun$ which approximately matches the mass range of observed galaxies 
corresponding to the data points. \label{fig:ssfr}}
\end{figure*}

At high $z$, the SFRs for the most massive galaxies appear similar among all the CR runs
(Note, these galaxies are absent in the UCR runs). The rates start to 
diverge below $z\sim 9$, with the CW model becoming the dominant one, followed by the VW and the NW, 
each lower by a factor of 2 -- 3 by $z=6$. For these galaxies, the SFR has reached 
$\sim 10^3\,\msun\,{\rm yr^{-1}}$, bringing them into the observed regime of LBGs at high-$z$ 
\citep[see also][]{Yajima.etal:15}. We also observe a similar trend in the intermediate mass 
bins, where SFR in the NW model start 
to decrease after $z\sim 8$, while models including galactic outflows continue to increase their SFRs.
By the end of the simulations, the CW model has the highest 
median SFR with $\sim 30-40\,\msun\,{\rm yr^{-1}}$ for $9.5<\log\,M_\star^{z=6}<10.5$, and
$\sim 600-800\,\msun\,{\rm yr^{-1}}$ for $10.5<\log\,M_\star^{z=6}<11.5$. 
The corresponding median SFRs in the NW run are $\sim 10\,\msun\,{\rm yr^{-1}}$ and 
$\sim 100\,\msun\,{\rm yr^{-1}}$, and the VW model ends up with intermediate values between the CW
and NW runs.

These results indicate that, even at high-$z$, galactic outflows do play an important  
role in controlling galaxy growth in overdense regions.  In the NW run, the absence of these outflows 
means the absence of a strong feedback which can delay or quench star formation. 
As a consequence, the peak of star formation in the NW run happens at 
higher redshifts, $z > 6$, as noted above, and thus the available gas supply for SF in galaxies is already 
consumed by $z=6$. Subsequent gas accretion could in principle increase SFRs again at later times in the 
NW run but, at $z=6$, 
we see that accretion has not yet been able to replenish the gas content in galaxies to maintain high SFRs. 
For this reason, SFRs are higher in CW and VW models at $z=6$ since the feedback from winds in those models 
acts to effectively delay the peak of star formation activity.

This is in agreement with \citet{Choi.Nagamine:11} and \citet{Angles.etal:14} which have analyzed 
galaxy evolution in average density regions, but lack massive galaxies at these redshifts. 
However, since galaxy evolution is accelerated in the overdense regions, the peak of SF activity 
happens at earlier times than in average density environment.
When comparing CW and VW models, results published in the literature show that 
the CW SFRs are generally the lowest ones, as measured in simulations at 
low $z\sim 3-4$ \citep[e.g.,][]{Choi.Nagamine:11}. However, the cosmic SFR evolution presented in 
\mbox{\citet{Choi.Nagamine:11}} (see their Fig. 5) shows that the difference between CW and VW SFRs 
decreases with 
increasing redshift at $z>6$, and, that these two models eventually reach similar SF levels at high-$z$. 

\subsection{Specific star formation rates}
\label{sec:ssfr}

For a more detailed picture of the SF efficiency in different runs, we 
investigate the specific star formation rate (sSFR), which represents SFR 
per unit mass, and essentially indicates the production rate of stellar mass  
per stellar mass. Figure~\ref{fig:ssfr} displays the median sSFR-stellar mass relation in 
galaxies at $z=6$ (left frame) and the sSFR evolution with redshift (right frame) in all runs. 
Observational data from \citet{Labbe.etal:13} ($z=8$), \citet{Duncan.etal:14} ($z=6$ and 7), 
\citet{Salmon.etal:15} ($z=6$) and \citet{Tasca.etal:15} ($z=5.5$) have been added to both panels. 
In the left panel, galaxies are binned by stellar mass with 0.3\,dex interval. 

We observe a nearly constant sSFR with $M_\star$ in the NW and VW models, with sSFR $\sim 1.5$ and 
$2.5\,\mathrm{Gyr}^{-1}$, respectively. At the same time, the 
CW model shows an overall positive correlation between sSFR and $M_\star$. 
This upward turn above the characteristic mass of $\sim 10^9\,\hmsun$ is in contradiction 
with observational trend shown on Figure~\ref{fig:ssfr}.

We also find that the sSFR in the CW run fall within the observed range, but the exsistense
of the peak in the mass range of $\sim 10^9-10^{10}\,\hmsun$ does not agree with the observations
of \citet{Salmon.etal:15}. On the other hand, for the NW run, the 
overall trend roughly agrees with the observed one, but the sSFR lies outside of the observed range. 
The best fit to observations is that of the VW model --- both the trend with $M_\star$ and the sSFR 
range agree, although the median observed fit is slightly shifted away from the model values.

\begin{figure*}
\epsscale{0.8}
\plottwo{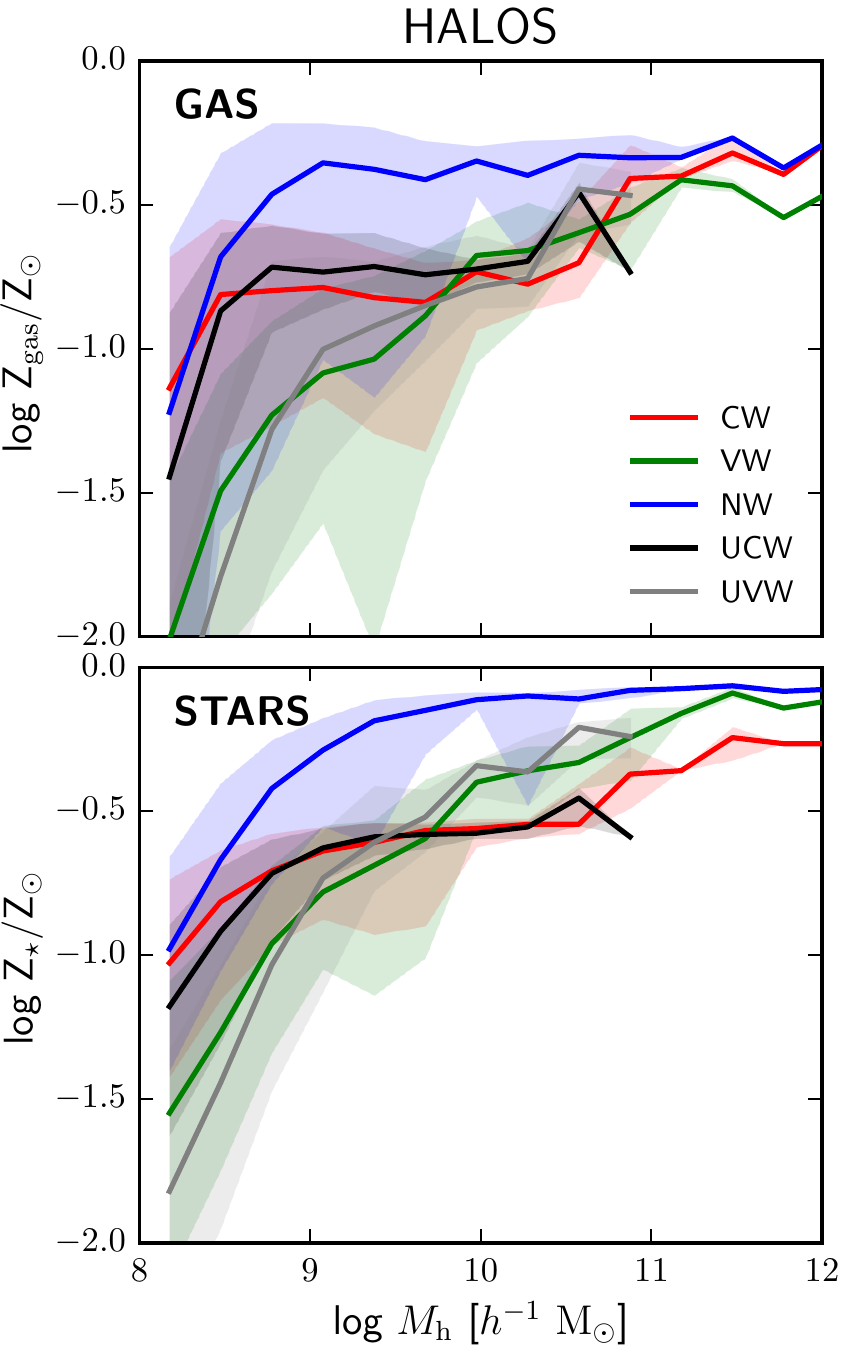}{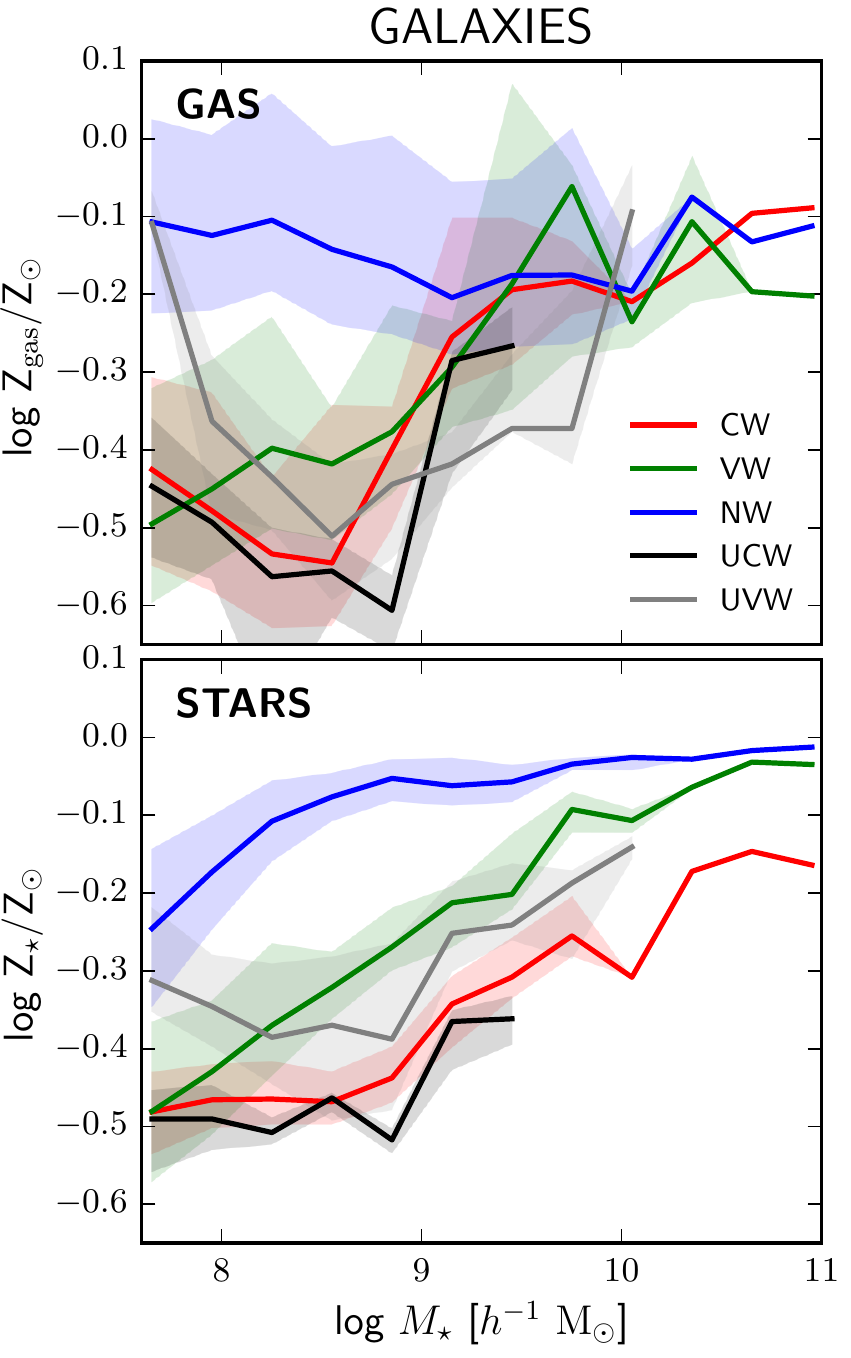}
\caption{
{\it Left:} Gas (top) and stellar (bottom) metallicities of halos (including gas
in galaxies) as a function of halo total mass at $z=6$ in all simulation runs.
{\it Right:} Gas (top) and stellar (bottom) metallicities of galaxies as 
a function of stellar mass at the same redshift. The solid lines indicate the median
values, while the colored regions correspond to the values between 
the 20 and 80 percentiles. Metallicity profiles of the IGM gas are shown 
in Fig.\,\ref{fig:TZ_profiles}. \label{fig:halogas_Mhalo_Z}}
\end{figure*}

The right-hand panel of Figure~\ref{fig:ssfr} shows the evolution of the sSFR with redshift. Solid lines 
correspond to the median trends for the whole galaxy population. The dashed lines show the 
evolution only for galaxies with stellar masses in the range $10^9-10^{9.5}\,\hmsun$, corresponding 
approximately to the estimated masses of objects targeted by observers. 
We find the median evolution of all galaxies 
to be dominated by the low and intermediate-mass objects because of a higher number of these galaxies compared 
to the massive ones. The median sSFR of all galaxies in the CW and UCW runs are 
in good agreement --- again underlying that gas content and SF in low-mass galaxies are not 
affected significantly by the overdensity. 

The sSFR in the NW run is the highest for $z\ga 7$, 
where it falls below the VW model. In all models except VW, the sSFR drops by almost 
an order of magnitude, from $\sim 10\,\mathrm{Gyr}^{-1}$ at $z=12$ to $\sim 1\,\mathrm{Gyr}^{-1}$ at $z=6$. 
The sSFR evolution in the VW case exhibits a trend similar to the one found 
for the gas fraction in low-mass galaxies (Fig.~\ref{fig:fgas_halos}). As we have argued in the 
previous section, This trend is caused by scaling of the feedback strength with the SFR 
in the VW run. At high-$z$, outflows quench the SF, which in turn decreases the feedback from 
the winds and allows an elevated SFR at later times. 
The median sSFR evolution for the whole galaxy samples 
does not reproduce correctly the observations for $z < 8$. The explanation of the mismatch lies in the 
dominant contribution from low-mass galaxies ($M_\star \la 10^9\,\hmsun$) to the global sSFR. 
When comparing the evolution for galaxies in the observed stellar mass bin (dashed lines), 
both CW and VW, as well as UCR, runs appear to be in a good agreement with the observed 
increase of sSFR with redshift. 

\subsection{Mass-metallicity relation}
\label{sec:M-Z_relation}

Next, we focus on the mass-metallicity relation in halos and galaxies. 
Figure~\ref{fig:halogas_Mhalo_Z} (left frames) shows the metallicity of gas and stars in halos 
as a function of the total halo mass at $z=6$.
The highest metallicities are found in the most massive halos because their star formation 
starts earlier and they exhibit a higher SFR compared to lower mass objects. 
We observe a rapid increase in the metallicity in the low-mass halos, and its leveling off at the 
intermediate/high mass end. 
The CW and VW runs reach close to solar metallicity for $M_{\rm h} \ga 10^{11}\,\hmsun$ by the end of the 
simulations. For smaller $M_{\rm h}$, the CW and VW models
display a slow decline in $Z$, which steepens dramatically below $M_{\rm h}\sim 10^{10}\,\hmsun$ for VW,  
and below $\sim 10^9\,\hmsun$ for the CW. This `knee' is moving gradually toward higher masses with 
redshift. 

While halos with $M_{\rm h}\sim 10^{11}-10^{12}\,\hmsun$ have a median $Z/Z_\odot\sim 0.6$ (gas) 
and $\sim 0.8$ (stars), the lower mass halos, with $\sim 10^8-10^9\,\hmsun$, show  
$Z/Z_\odot\sim 0.01-0.03$ (gas) and $\sim 0.03$ (stars), respectively. Note that the NW model 
achieves the highest metallicity in all 
$M_{\rm h}$, for both gas and stars. Furthermore, compared with the halo gas metallicities, stellar 
metallicities display higher median $Z$ --- this means that the gas metallicity is diluted by the
cold accretion from the IGM, in agreement with \citet{Romano-Diaz.etal:14}.  

Galaxy metallicities in the gaseous and stellar components are shown in Figure~\ref{fig:halogas_Mhalo_Z} 
(right frames).  
As expected, the NW model metallicity exceeds substantially those of CW and VW in both components,
due to the metals being locked in galaxies and their immediate vicinity. A possible exception
may be the highest mass bin for the ISM, where CW model shows marginally higher metallicity.
The bimodal behavior
observed for the halos, can be seen here as well. Stars also remain more metal-rich than
the gas, again in tandem with the halos. Future observations will constraint these models. 
We point out that, in Figure~\ref{fig:halogas_Mhalo_Z}, we have only computed the average metallicity 
inside individual galaxies and ignored the scatter in metallicity in a given galaxy.

\section{Results: Effect of winds on environment}
\label{sec:wind_envir}

\begin{figure*}
\plottwo{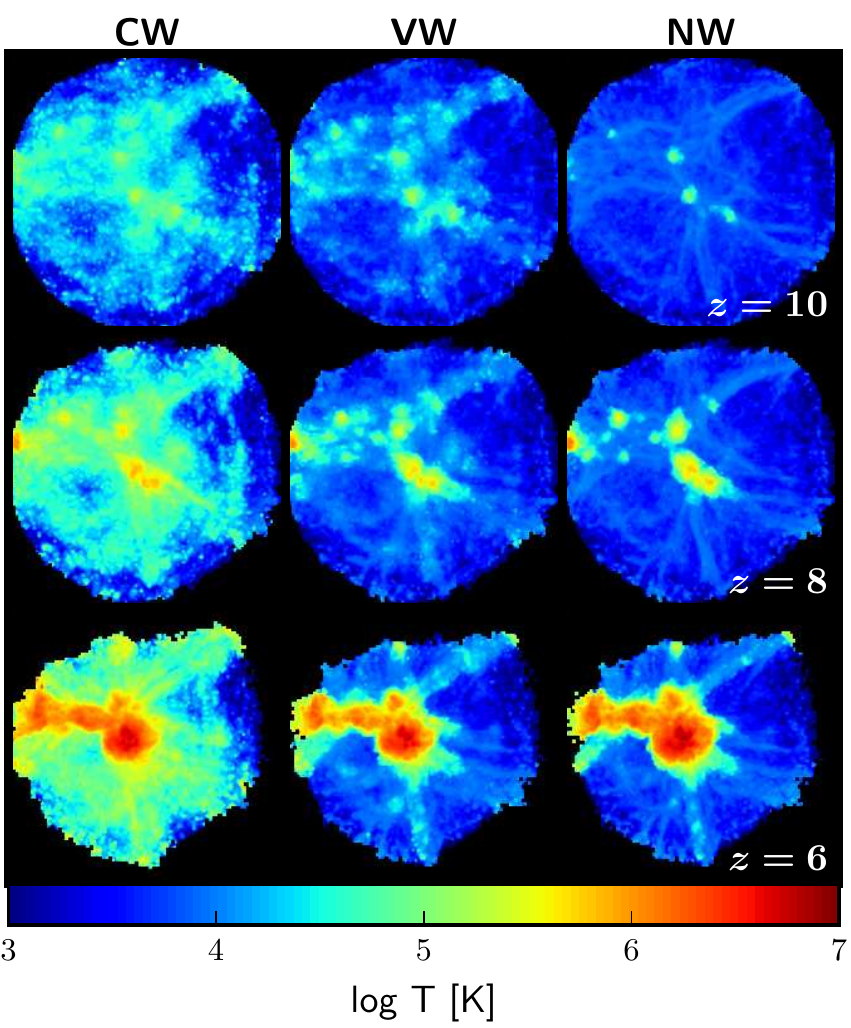}{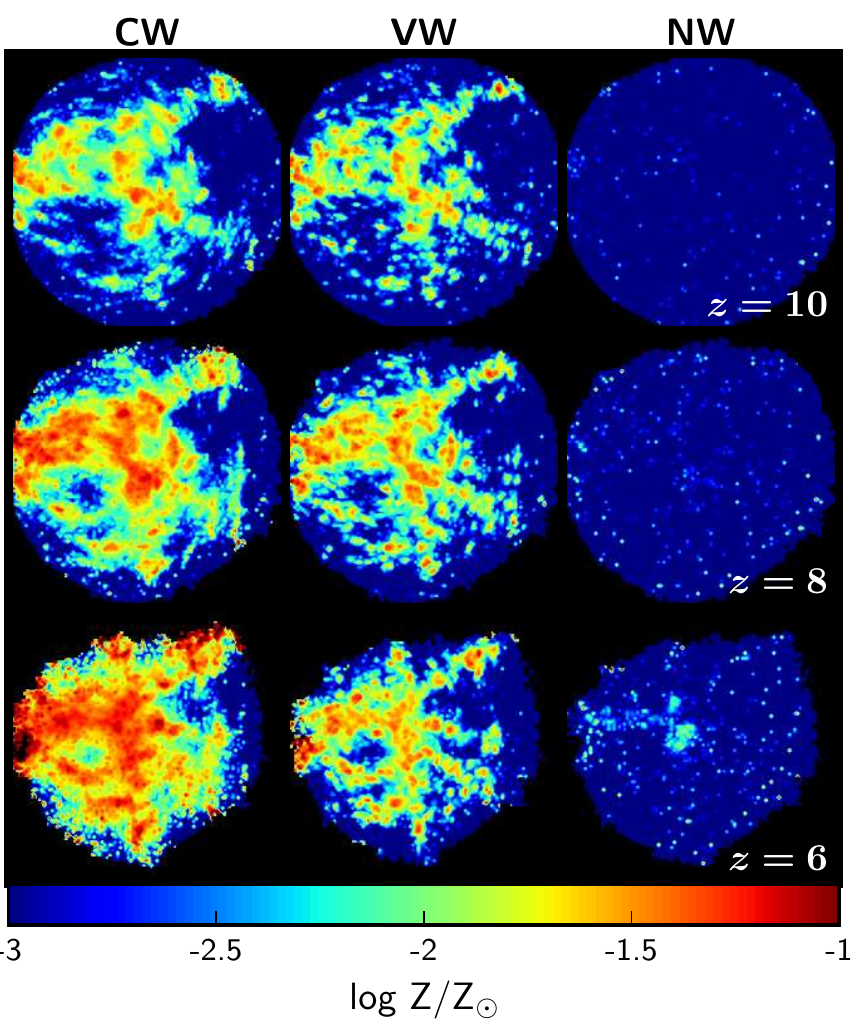}
\caption{
Temperature (left) and metallicity (right) maps of the IGM gas (excluding the gas inside 
halos) within the central $2.7\,h^{-3}{\rm Mpc}^3$ inner region, at $z=10$, 8 and 
6 (top to bottom) for the three CR models (CW, VW and NW). \label{fig:tmaps}}
\end{figure*}

\begin{figure}
\epsscale{0.8}
\plotone{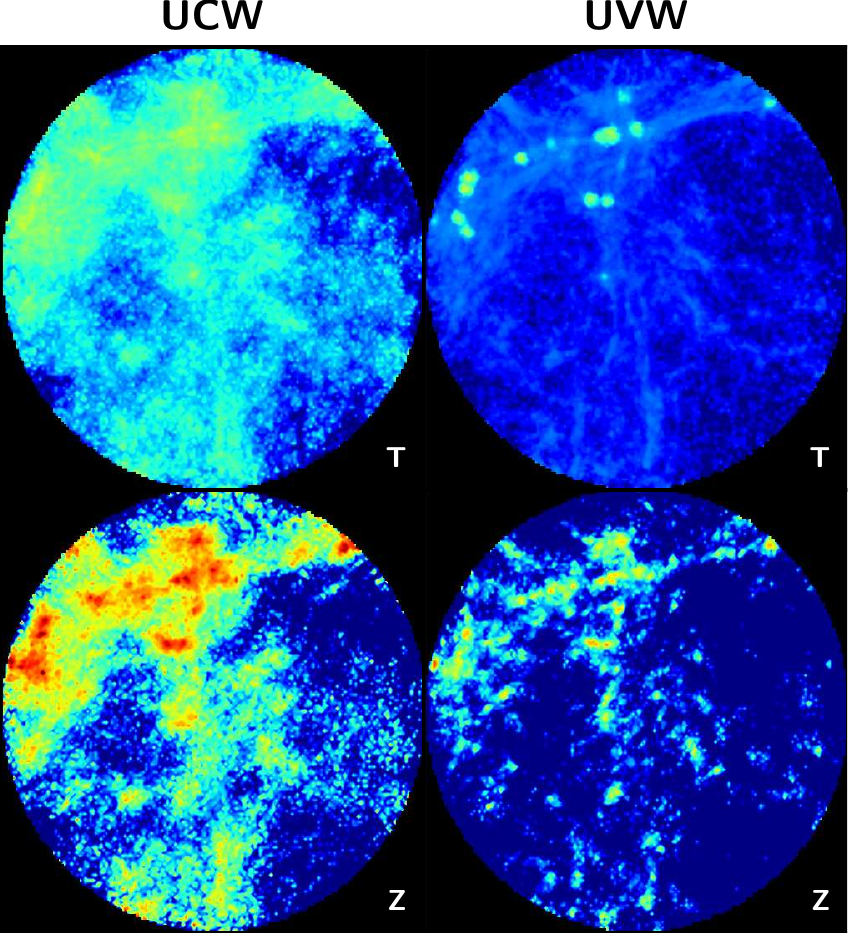}
\caption{
Temperature (left) and metallicity (right) maps of the IGM in the UCR 
runs at $z=6$ within the central $4.0\,h^{-3}{\rm Mpc}^3$ inner region. 
The color scales are the same as in Fig.~\ref{fig:tmaps}. \label{fig:TZmaps_UCR}}
\end{figure}

In this section we focus on the effects of winds on the thermal evolution and metal enrichment 
in the circumgalactic and intergalactic medium. Wind effects on the ISM has been discussed
in section\,\ref{sec:fractions_env}.

\subsection{Temperature and metallicity maps}
\label{sec:igm}

The temperature maps of the IGM gas, i.e., gas outside galaxies and their host halos, in the 
inner central region of the simulation box clearly display the growing difference
between the wind models with redshift. Figure~\ref{fig:tmaps} (left frames) underlines  
profound differences existing between the CR runs already at $z\sim 10$. The NW run lacks gas with 
$T\ga {\rm few}\times 10^4$\,K, except for a few spots near the central density peak 
corresponding to the growing massive halo. The warmest gas can be found in the
large-scale filaments, and it roughly delineates them in all CR models. 
The ${\rm few}\times 10^4$\,K gas is more closely associated with large-scale structures in the VW case, 
because the feedback from outflows is lower compared to the CW run, where this gas ``spills out'' of 
the filaments. The rest of the volume hosts gas with  $T\la 10^4$\,K.

The redshift evolution of these maps shows gradually heating and cooling regions,
of ${\rm few}\times 10^5$\,K gas at $z=8$, and ${\rm few}\times 10^6$\,K gas at $z=6$. By the end of
the run, 
the massive central halo is surrounded by a hot bubble of $\sim 10^7$\,K gas in CW and VW wind models. 
Away from the most massive halo, the filament temperature drops, most profoundly in the NW model. 
By $z\sim 6$, the CW gas in nearly the whole volume is heated up to $10^5-10^7$\,K. 
In this model, the $10^5-10^6$\,K  gas is present in the main filament, and $\sim 10^5$\,K gas in other 
filaments. In VW and NW models, the gas outside the filaments remains cold. 

Hence, at higher redshifts, $z\sim 10$, we observe increasing temperatures
of the IGM gas, along the sequence from NW, to VW, and to CW. 
With decreasing $z$, the central bubble heats up, and so 
do the filaments. The outflows in the CW model are capable of heating up the IGM, while the VW outflows
have a dramatically lower impact on the IGM temperature at these redshifts, 
but nevertheless heat it up toward $z\sim 6$. 

Figure~\ref{fig:tmaps} (right frames) shows maps of the metallicity distribution 
in the IGM gas at similar redshifts, for all the CR models.  
At $z=6$, the IGM peak metallicity has reached $\sim 0.1 Z_\odot$ inside the central hot bubble and along
the main filament corresponding to the regions surrounding the massive central structure 
in the CW and VW models. The peak metallicity in the NW is a factor
of a few lower even though the SFRs are higher in this model at high-$z$. 
The extended region in the CW model shows a negative metallicity gradient away from the central bubble
and the main filament. Still, most of the volume is polluted by metals and has 
$Z\ga 10^{-2} Z_\odot$. The VW exhibits a much lower metallicity outside the filaments,
$Z\la 10^{-3} Z_\odot$, whereas the NW IGM remains mostly pristine.
The reason for this is the absence of outflows, which causes metals 
to be locked in galaxies and their halos. In comparison, 
solar metallicities have been reached only within galaxies of the NW model 
(see Fig.~\ref{fig:halogas_Mhalo_Z}).

Figure~\ref{fig:TZmaps_UCR} displays the temperature and metallicity maps of the 
IGM in the UCR runs at $z=6$, and can be compared directly with the CW and VW maps in 
Figure~\ref{fig:tmaps}, 
as both figures use the same color scales. Since the UCR runs represent an average density region 
and lacks massive halos (e.g., Fig.~\ref{fig:halo_gal_mfs}), both the peak and average IGM temperatures 
are dramatically lower than in the CR runs. The peak metallicity in the IGM and its average are also lower 
compared to the CR runs. Thus the impact of the overdensity is readily visible from these two figures --- 
the average density universe remains relatively cool at $z\sim 6$.

\subsection{Temperature and metallicity profiles}
\label{sec:igmprofs}

\begin{figure}
\plotone{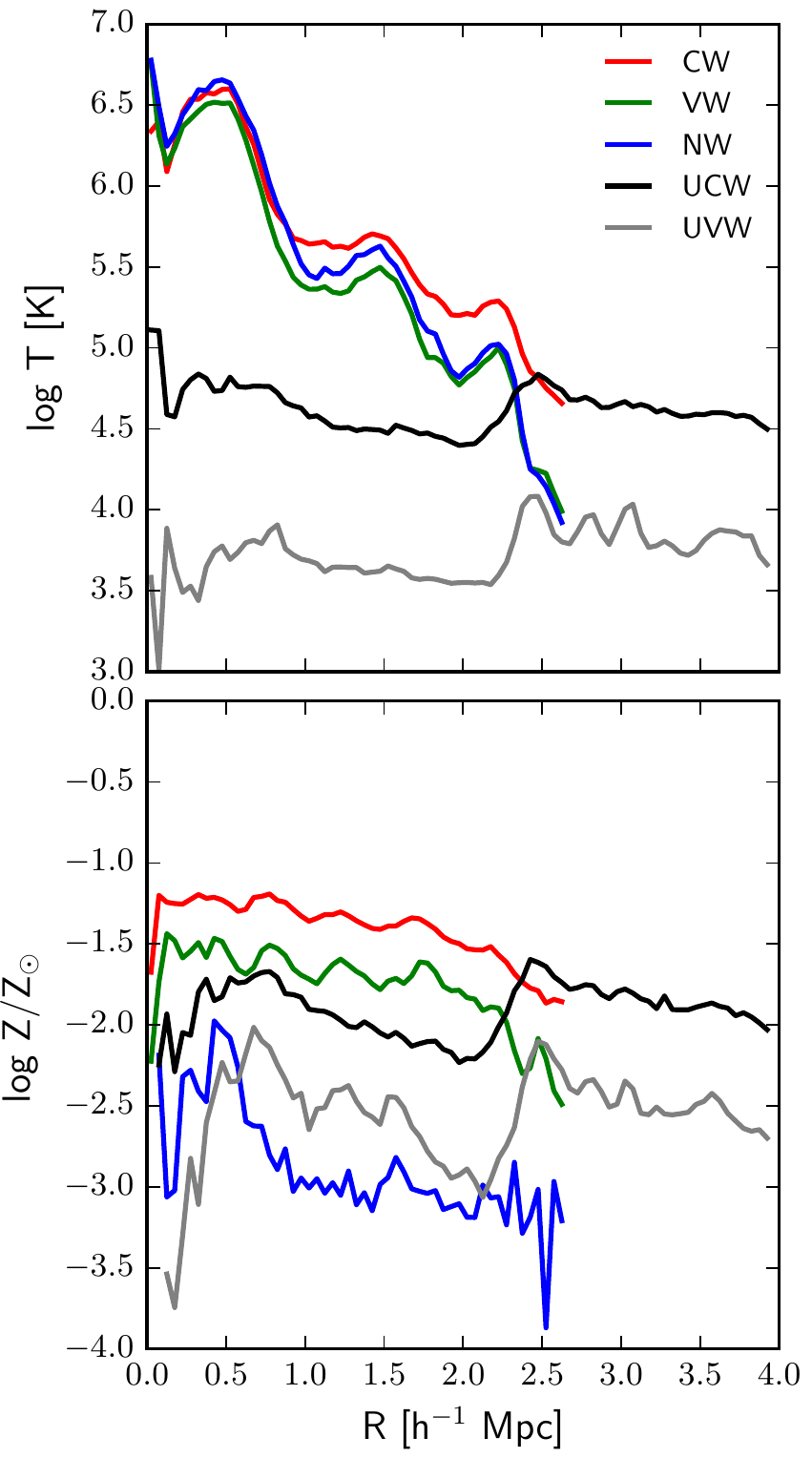}
\caption{
Temperature (top panel) and metallicity (bottom panel) 
radial profiles of the IGM gas at $z=6$ calculated from the 
center of the computational domain to the radius $R_{\rm inner}$ 
of the inner high-resolution region 
(see table \ref{table:sim_properties}). \label{fig:TZ_profiles}}
\end{figure}

A more quantitative comparison between the different runs 
on the state of the IGM at $z=6$ is displayed in 
Figure~\ref{fig:TZ_profiles} which shows the spherically-averaged 
IGM temperature (top) and metallicity (bottom) profiles 
from the center of the computational box to the 
radius $R_{\rm inner}$  of the inner high-resolution region, where 
$R_{\rm inner} = 2.7\,h^{-1} \mathrm{Mpc}$ and $R_{\rm inner} = 4\,h^{-1} \mathrm{Mpc}$ 
for the CR and UCR models respectively. The profiles are computed by 
averaging the temperature and metallicity of the IGM gas on 
spherical shells of thickness $50\,h^{-1} \mathrm{kpc}$. 

Both the temperature and metallicity profiles agree with the qualitative results 
from Figures~\ref{fig:tmaps} and \ref{fig:TZmaps_UCR}. 
The CR runs have similar IGM temperature within the 
central $1\,h^{-1} \mathrm{Mpc}$, corresponding approximately to the 
size of the imposed constraint which is visible as a hot bubble on Figure~\ref{fig:tmaps}. 
In this central region, the IGM temperature reaches $T\sim 10^7$\,K which is consistent 
with the virial temperature for a halo of mass $\sim 10^{12}\,\msun$, 
corresponding to the mass of the central halo that forms in the overdense region. 
On the other hand, the IGM temperature in the UCR runs 
remains constant with radius at much lower values, 
around $\sim 10^{3.5}$ - $10^4$\,K and $\sim 10^{4.5}$ - $10^5$\,K 
for the UVW and UCW models respectively. Thus, it appears that shock-heating 
caused by the presence of the massive structure in the overdense region is largely responsible 
for the elevated IGM temperature within $1\,h^{-1} \mathrm{Mpc}$ in the CR runs. 
This also explains why the CR runs show similar IGM temperature in the central region as 
the differences in temperature due to the different wind models (which are much smaller) 
get washed out.

The metallicity profiles (Figure~\ref{fig:TZ_profiles}, bottom) are 
in general much flatter than the temperature ones. The lowest 
metallicities are seen in the NW run, followed by the VW and CW models, 
in agreement with Figure~\ref{fig:tmaps}. The UCW 
IGM metallicity is smaller than both CR models 
including winds (CW and VW) within $\sim 2\,h^{-1} \mathrm{Mpc}$ and then rises on 
larger scales to reach levels comparable with the CW run. This increase is associated 
with the presence of a massive filament in the UCR runs (Figure~\ref{fig:TZmaps_UCR}), as 
a similar rise can be seen in the UCW temperature profile 
at the same distance from the center.

\subsection{Gas phases at $z=6$}
\label{sec:gas_z=6}

\begin{figure*}
\epsscale{0.95}
\plotone{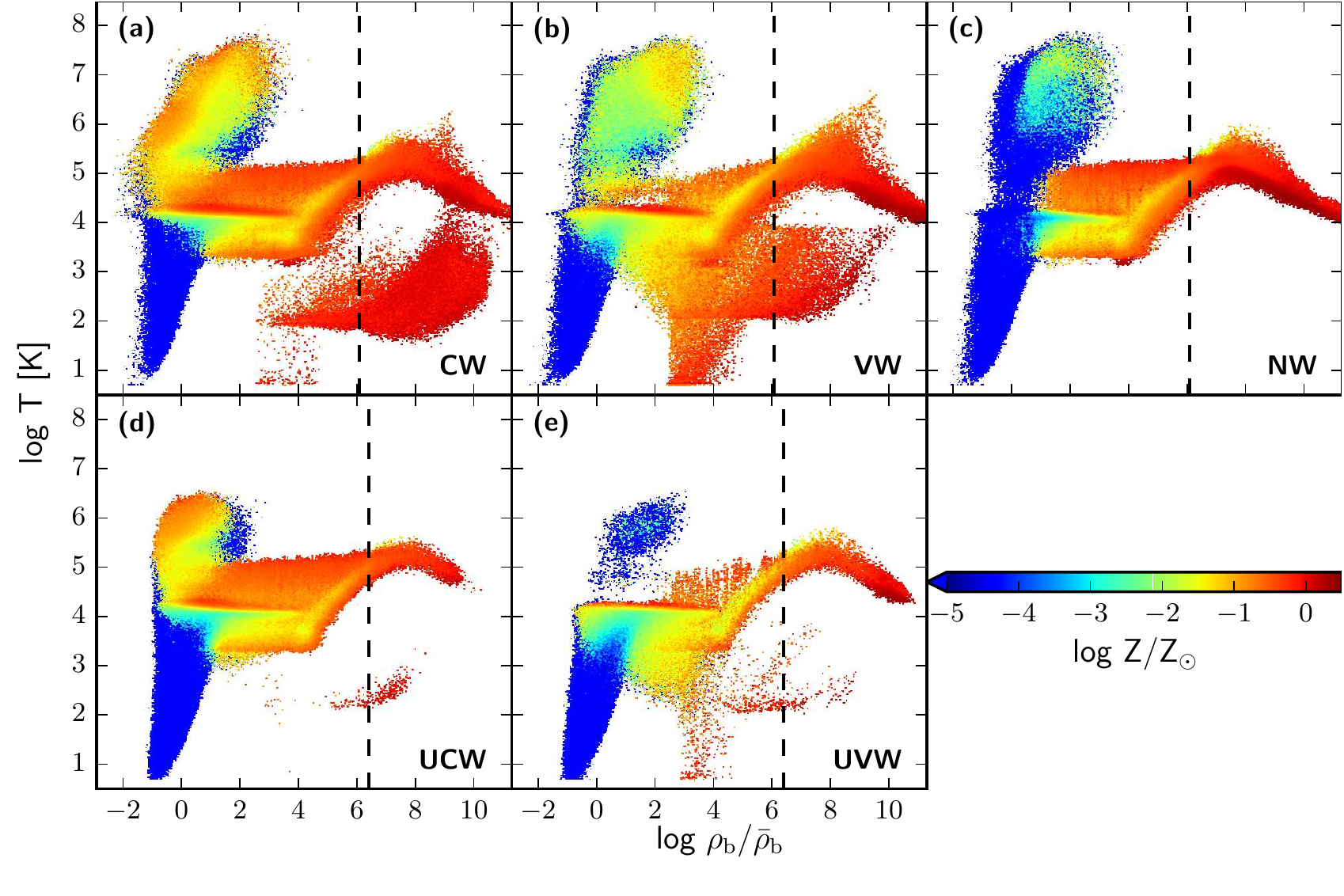}
\caption{
Distributions of all gas particles found in the high-resolution region
in the temperature-density plane at $z=6$ in all the simulation runs:  
{\it (a)} CW, {\it (b)} VW, {\it (c)} NW, {\it (d)} UCW, and {\it (e)} UVW.
Gas particles are colored based on their metallicity as indicated by the color scale on the right.
The density is normalized by the average baryon density in the high-resolution computational box
which is different between CR and UCR runs. For this reason, the vertical dashed line 
corresponding to the star formation threshold density $n_{\rm crit}^{\rm SF}$ is offset in the 
bottom panels (UCR runs) compared to the one in the top panels (CR runs).
\label{fig:Z-T-rho_maps}}
\end{figure*}

A supplementary global view on the temperature and metallicity of the gas can be taken from
Figure~\ref{fig:Z-T-rho_maps}, where we show the gas phase diagrams in the 
temperature-density plane at $z=6$ for all gas particles found in the high-resolution region.
Gas particles are also colored based on their metallicity. 
The densities are now normalized by the average baryon density in the 
high-resolution region, as opposed to using the universal value like 
we did in section \ref{sec:fractions_env}. This normalization is thus different 
between CR and UCR runs since the average density is higher in the overdense region. 
For this reason, the vertical dashed line, which separates the non-starforming (to its left)
from starforming (to its right) gas, is slightly offset in the bottom panels (UCR) compared 
to the one in the top panels (CR).
We observe the highest metallicities, around solar values, in the
starforming gas which is confined deep within galaxies. One can easily identify the wind gas in 
the lower right part of the diagrams, as it is present in the VW, CW, UCW and UVW cases, 
and is metal-rich, but not in the NW case. 

The fraction of low-density gas polluted with metals, 
(green and red) compared to pristine (blue) and the highest metallicities attained there, depend 
strongly on the wind model. The metal-rich, low-density gas originates in the galactic outflows, and
heats-up by the shock-stopping of these high-speed outflows. This gas lies at temperatures of
$\sim 10^5-10^7$\,K gas and 
densities ${\rm log}\,\rho_{\rm b}/\bar{\rho}_{\rm b} \la 4$. Here the baryon density 
$\rho_{\rm b}$ is normalized by the average density in the high-resolution box.
We note that a large amount of the $\sim 10^4$\,K starforming gas found in 
our models at high densities is due to the delayed star formation algorithm, the {\it Pressure} model,
used (section 2.1), and because of the absence of H$_2$ cooling in our simulations.
The amount of the heated gas increases substantially from the NW, to VW and CW runs.
The hot and low density pristine gas at $T \ga 3\times 10^6$\,K in the CR runs consists mostly 
of virially shock-heated material infalling onto the most massive halo, and, as such, is not observed in the
UCR models.
The amount of intermediate metallicity, $Z/Z_\odot\sim 10^{-1}$, gas at $T\ga 10^6$\,K, 
is steadily increasing in all CR models.

\begin{figure}
\epsscale{1.1}
\plotone{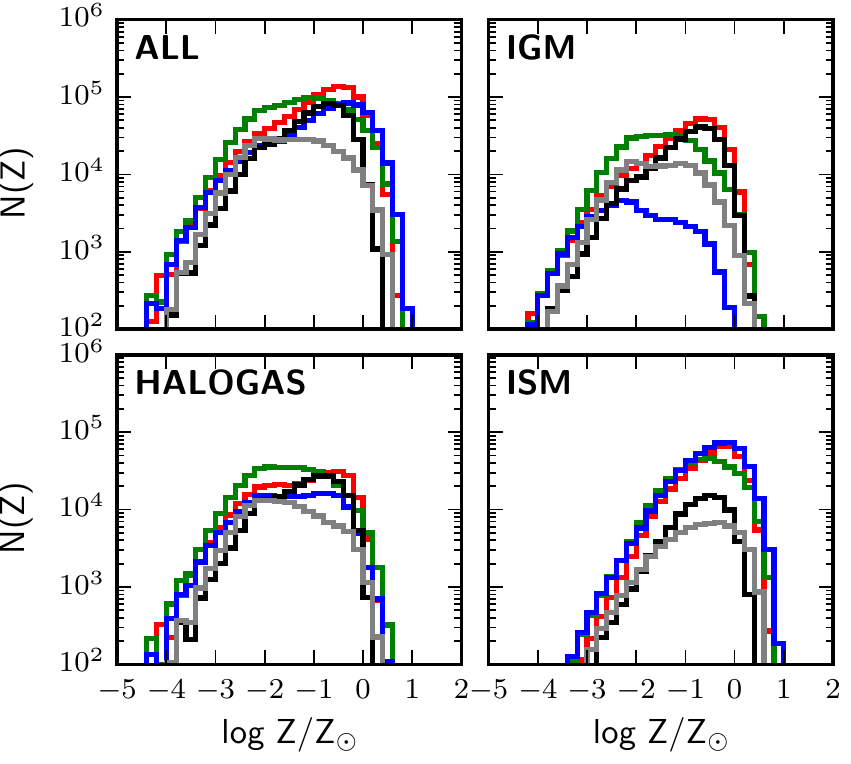}
\caption{Metallicity distributions for different gas phases at $z=6$ in
  the high-resolution region: All gas particles (top left), IGM (top right),
  halo gas (bottom left) and ISM (bottom right) for all the simulation runs: CW (red), 
  VW (green), NW (blue), UCW (black) and UVW (grey). 
  \label{fig:metal_distrib}}
\end{figure} 

\begin{figure*}
\epsscale{0.95}
\plotone{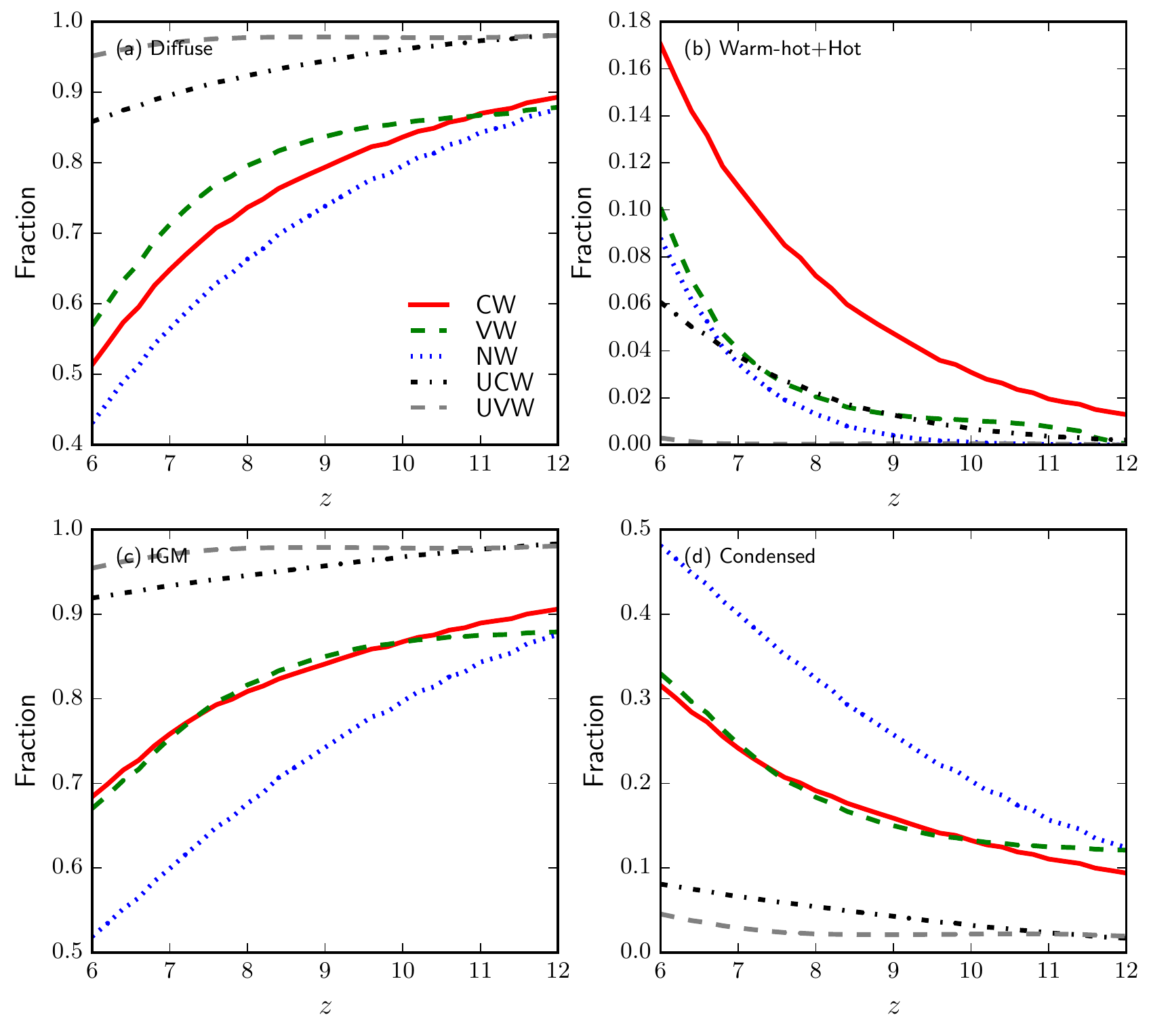}
\caption{Evolution of the mass fractions of the four different baryon phases (hot, warm-hot, 
diffuse, and condensed --- see section\,\ref{sec:gas_evolution_mass} for definitions) for the CR 
and UCR simulations (see the text for definitions).
Panel (b) shows the addition of the two phases, `warm-hot $+$ hot', and panel (c)
shows the total mass in the IGM, i.e., `diffuse $+$ warm-hot $+$ hot'. 
Most of the UCR baryons are in the `diffuse' phase, while in the CW the majority is found in
the `warm-hot' phase. The `condensed' phase is also more populated in the CR runs, 
especially for the NW model. \label{fig:four_gas_fractions}}
\end{figure*}

Figure~\ref{fig:metal_distrib} provides the metallicity distribution 
for different wind models in various gas phases --- the IGM, 
halo gas, and the ISM, as well as the total distribution for gas particles
in the high-resolution region.
The ISM includes all the gas inside galaxies, the halo gas includes the gas 
within halos but outside galaxies, and the IGM includes the gas outside the halos. In the IGM
panel, we observe that the CW models for CR and UCR exhibit the highest metallicities,
basically indistinguishable from each other, with maxima located at log\,$(Z/Z_\odot)\sim (-0.8)-
(-0.7)$. The least efficient injection of metals happens in the 
NW model (the maximum at log\,$(Z/Z_\odot)\sim -2.2$), while VW occupies broadly the region 
between the CW and NW models. It displays a nearly flat top at log\,$(Z/Z_\odot)\sim (-2)-(-1)$. On 
the other extreme, the metal distributions in the ISM are basically identical, and the maxima are 
found between log\,$(Z/Z_\odot)\sim (-1)-0$. The metal distributions in the halo gas are flat-top
for NW and VW models, and show maxima at log\,$(Z/Z_\odot)\sim -1$ for the UCW, and $-0.7$ 
for CW. From Figure~\ref{fig:metal_distrib}, we conclude that the ISM is polluted by metals in all
models. Moreover, there is a large difference in expelling the metals to the halo
and the IGM. The CW and UCW models appear to be the most efficient in this process in 
a given environment. 

\subsection{Gas phases evolution}
\label{sec:gas_evolution_mass}

To take an alternative view at the gas evolution, we subdivide the gas phases in a different way. 
In this section, we alter definitions for the IGM and other gas phases used
in the previous section, in order to facilitate the comparison with previous works.
Figure~\ref{fig:four_gas_fractions} compares the evolution of various gas phases, namely, 
diffuse, warm-hot-hot, IGM and condensed gas. These phases are defined based on the temperature 
of the gas and the density fluctuation $\delta = \rho_{\rm b}/{\bar{\rho}_{\rm b}}-1$. 
For this purpose, we employ the definitions from \citet{Dave.etal:99} and \citet{Choi.Nagamine:11}. 
The `diffuse' gas is defined by 
$T < 10^5$\,K and $\delta < 1000$, the `hot' gas by $T > 10^7$\,K, and the `warm-hot' gas by 
$10^5 < T < 10^7$\,K. The `condensed' phase consists of non-starforming gas with $T < 10^5$\,K and
$\delta > 1000$, and of a starforming gas as well as stars.

We observe that the mass fraction of the `condensed' phase is increasing with time for all models, 
especially for the NW, while the IGM (non-starforming) fraction decreases. The fraction of a `diffuse' 
filamentary gas is decreasing faster for the CW model than for the VW one. 
At the same time, both the UCW and UVW diffuse gas fractions remain very high at all $z$, while the condensed 
fraction stays very low. 
While in the warm-hot-hot phase the NW and VW evolution are similar, in the condensed and IGM
phases it is the VW and CW runs which evolve in tandem. 

While the mass fractions in our UCRs and \citet{Choi.Nagamine:11} models appear to be
reasonably similar, the CR models differ substantially, both from the UCR cases and among themselves.
This difference between the CR and the UCR models is expected, because of the overdensity 
present in the CR runs. But we also observe that the NW model exhibits a much higher fraction
of cold gas, as it evolves toward $z\sim 6$, and consequently, exhibits smaller fraction of IGM gas. 
The mass fraction of the condensed phase, \emph{which also includes stars}, in the NW run 
reaches $\sim 50\%$, while for the VW and CW models it is only $\sim 32\%$. 
The diffuse component shows that the fraction drops to $\sim 43\%$ in NW, to $\sim 51\%$ in CW and 
to only $\sim 57\%$ in VW. 

\section{Discussion and Conclusions}
\label{sec:discuss}

We have used high-resolution cosmological zoom-in simulations of average and overdense 
regions in the universe in order to follow their halo and galaxy population evolution at 
$z\ga 6$. Specifically,
we compared this evolution applying different galactic wind models in order to analyze the effects of
outflows on: 

\begin{itemize}

\item galaxy evolution, including their environment. 

\item thermodynamic state of the IGM.

\item the spread of metals in the ISM, DM halos and the IGM. 

\end{itemize}

We have tested 
three different prescriptions for galactic winds: a constant-velocity wind (CW) model from 
\citet{Springel.Hernquist:03}, a variable-velocity wind (VW) model based on the 
\citet{Choi.Nagamine:11} method and a model without outflows (NW). 
Each model has been followed down to $z=6$. In order to get an insight into 
environmental dependence of galaxy evolution, the CW prescription 
for galactic outflows has been applied to both a constrained (CR) and unconstrained (UCR)
simulations, representing overdense and average regions in the universe.

The most general conclusions which emerge from this work are that, 
regardless of the applied wind prescription, (1) galaxies in the overdense region
evolve much faster than their average-region counterparts \citep[e.g., Figures\,\ref{fig:frames} 
\& \ref{fig:halo_gal_mfs}, see also][]{Romano-Diaz.etal:14}, and (2) the  low-mass end of the
galaxy mass function
is shallower in the overdense regions --- an important issue to be addressed elsewhere. This is 
a direct corollary 
of the imposed constraint which has been designed to collapse by $z\sim 6$, based on the 
top-hat model. As a consequence, we observe an accelerated 
evolution of its environment within the correlation length 
of the constrained halo \citep[e.g.,][]{vandeWeygaert.Bertschinger:96}. 
DM halos in the vicinity of the constraint experience
a faster evolution compared to their average-density counterparts, inducing earlier 
baryon collapse, setting the
stage for star formation. We find that the galaxy population in such regions is
much more evolved by the end of simulation. This is indicated not only by the presence of more 
massive galaxies (up to $1-2\times 10^{11}\,\hmsun$ in stellar mass) which are absent in the 
average density region (where  $M_\star \la 10^{9.5}\,\hmsun$), but also by a substantially
larger number of galaxies over the mass range considered here.

Finally, we observe that galactic winds have a substantial effect on galactic environment,
which includes modifying its thermodynamic state and metallicity. While CWs show little
dependence on the overdensity, the corollaries of VWs are more diverse. This is clearly related
to the amount of energy deposited by the VWs in DM halos and IGM, which varies with galaxy
properties, and, in general, is smaller than that deposited by the CWs. In other words, the
VWs appear less ``destructive'' than their constant velocity counterparts, and more fine-tuned
to the galaxy growth mode in the form of cold accretion flows --- their overall effect differs
between overdense and average regions. On the other hand, the CWs
weaken the cosmological filament flows around galaxies, irrespective
of surrounding densities.

Galactic winds, as a form of feedback, serve as main agents in distributing recycled, 
metal-enriched gas, which is used to form new stars.  
The effect of winds onto the evolution of our modeled galaxies can be most clearly
observed in the galaxy mass function (Fig.\,\ref{fig:halo_gal_mfs}, bottom panel), gas fractions 
(Fig.\,\ref{fig:fgas_halos}) and star formation rates (Fig.\,\ref{fig:sfh}). 
The IGM properties can be affected by galactic winds, so comparing the 
observed IGM properties would provide the clearest justification for the wind models.
However, the IGM at $z\ga 6$ is not fully reionized, and the conventional IGM observation 
technique, e.g. quasar absorption line measurement, cannot be used in this epoch.
This limits the IGM observation resources that can constraint our wind models.
However, the pre-reionized $\rm H\,I$ status at $z\ga 6$ provides an interesting 
implication that can discriminate between our CW, VW and NW models, as discussed below.

The model with no-wind feedback 
produces the largest number of galaxies because the gas is hardly
expelled from their DM halos. In this case, the feedback is limited to the thermal feedback by the SN, 
and, therefore, contributes to the gas and galaxy survivals within the DM substructure. 
Most of this gas remains locked up within their DM halos, cools down,
and is accreted by the galaxies \citep[e.g.,][]{Keres.etal:05,Dekel.Birnboim:06,Romano-Diaz.etal:14}. 
As a consequence, these
galaxies become very metal-rich early-on (Fig.\,\ref{fig:halogas_Mhalo_Z}), 
enhance their SFRs (Fig.\,\ref{fig:fgas_halos}), consume their gas very 
quickly (Fig.\,\ref{fig:fgas_halos}), and
become gas-poor and very compact already by $z\sim 6$.
An overall steady decay in the gas fraction within the NW galaxies is correlated with
their SFRs (Fig.\,\ref{fig:sfh}).
The smallest effect is observed in the lowest mass-bin, i.e., $7.5 < 
{\rm log}\,M_*^{z=6} < 8.5$. This happens
because such objects are less prone to the feedback effects due to
their respective, already very low SFRs.

Our conclusions regarding the NW model pertain mostly to the decline in the gas
fraction in intermediate-mass and massive galaxies, and consequently the
decline in their respective SFRs. This implies 
reddening of such galaxies due to aging of their stellar populations at relatively 
high-$z$ (e.g., $z\ga 3$). However, such a trend seems to be in
contradiction with the observed rise in the cosmic
SFR which reaches its peak at $z\sim 2-3$ (see \citet{Madau.Dickinson:14} for the latest
review), although the latter SFR is averaged over all possible environments. 
Furthermore, this model is in disagreement with the measured
SFR at $z\sim 6-8$ (Fig.\,\ref{fig:ssfr}) due to an early gas depletion, and deviates
most from the empirical sSFR -- $M_*$ median shown in Figure\,\ref{fig:ssfr}.

\begin{figure}
\includegraphics[width=\columnwidth,height=\columnwidth]{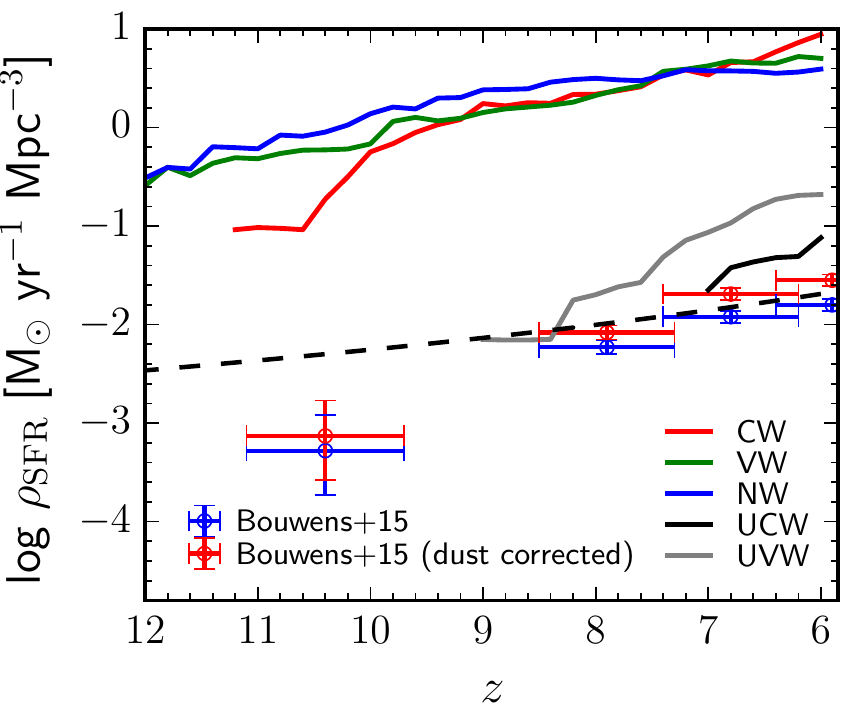}
\caption{Evolution of star formation density with respect to the total computational box for all 
galaxies that have $M_*\ga 10^9\,\hmsun$ at each redshift.
This is different from Figure\,\ref{fig:sfh} which provides star formation histories
for galaxies chosen at $z=6$. The dashed line is the best fitting line from 
\citet{Madau.Dickinson:14}, and is based on UV and IR data. The
observational points are from \citet{Bouwens.etal:15} --- blue points are uncorrected for
dust, while red points have been corrected for the dust extinction. \label{fig:global_sfr}}
\end{figure}

For a more direct comparison with observations of high-$z$ galaxies, we have calculated
the global star formation density in our computational box for galaxies with 
$M_*\ga 10^9\,\hmsun$ (Fig.\,\ref{fig:global_sfr}).
We include all the wind models in the overdense and normal regions. Furthermore, we have added
the best-fitting curve from \citet{Madau.Dickinson:14}, given by their Eq.\,15 (see also
their Fig.\,9), as well as available observational points from \citet{Bouwens.etal:15}.

A number of conclusions can be made based on Figure\,\ref{fig:global_sfr}. First, the 
overdense models lie well above the normal density models. 
However, their slope is in good agreement with the observational data points,
while their amplitude is above that of the data, as expected 
\citep{Romano-Diaz.etal:14,Yajima.etal:15}.
Second, an extrapolation of the trend from the UCW and UVW models to
higher $z$ show that their respective
curves have a slope which agrees well with the observational
points at $z\sim 7 - 10$, although it is somewhat steeper for $6\la z < 7$.
Lastly, this agreement is better than the one provided by the
best-fitting curve (dashed line) at $z\ga 8$. Hence, our UCR curves, although
lying above the dashed curve at $z\sim 6-7$, provide a better match to
observations at higher redshifts. Note, that latest corrections to the
observational points (both dust corrected and uncorrected) move up
steadily, e.g., when comparing them in \citet{Bouwens.etal:14} with 
\citet{Bouwens.etal:15}. Given the uncertainties of the current observations, our
small computational box, and the absence of any fine-tuning or
calibration in our simulations, the match between our UCR models and
observations is acceptable, which is important when one discusses the
overdense regions.

An immediate consequence of the NW model and the locked-up gas content
within their respective DM halos at $z\ga 6$ is related to the
reionization process of the universe and the metal
pollution/enrichment of the IGM. If the metal-rich gas is locked within
the DM halos and is highly concentrated, the dust content
within those halos and galaxies will be relatively high, making
difficult for the UV photons to escape and ionize their
surroundings. The escape fraction of photons will be lower than
the expected fraction of $\sim 0.3$ required to maintain the universe fully-ionized
by $z\sim 6$ \citep{Finkelstein.etal:12,Kashikawa.etal:11},
resulting in a delay in the process of reionization.

In comparison, the CW model points to a relatively high temperature IGM gas 
(Fig.\,\ref{fig:tmaps}) that extends well beyond the halo boundaries, deep into low
density regions, where most of the IGM
has reached $T\ga 2-3\times 10^4$K by $z\sim 8$.
This temperature can collisionally ionize the $\rm H\,I$, even in the absence of UV photons. 
This is in contradiction with the currently accepted scenario for reionization in which 
the neutral gas in the IGM is photoionized by the UV photons coming from galaxies and quasars. 
By $z\sim 6$, the CW model shows an overheated IGM gas, and we observe the formation of 
a hot bubble constituted 
of a shock-heated gas with a temperature up to $\sim 10^7$\,K 
in the central region. This CW model gas, potentially detectable 
as an X-ray bubble in this overdense environment, 
is absent in its UCW counterpart (Fig.\,\ref{fig:four_gas_fractions}). 
Importantly, both CW overdense model and the UCW model exhibit large volume-filling
ionized gas with $T\ga 10^{4.5}$K, which contradicts the re-ionization constraint.

Both NW and CW models exhibit a minimal dependence on the host galaxy properties. 
However, galactic outflows are expected to be driven by various mechanisms, 
including radiation pressure, SN and stellar winds, and photo-heating of H\,II regions 
\citep[e.g.,][]{Choi.Nagamine:11,Hopkins.etal:12,Agertz.Kravtsov:15}. Significance of each 
process depends 
on the galaxy properties and one should consider all mechanisms in the galaxy formation 
simulations to properly account for the cumulative effects.  Within this framework, the 
galactic outflows are hardly described by a simple scaling relationship. Consequently, we 
turn to the effects of a more complex type of models, the VW (section\,\ref{sec:winds}).
 
In the VW model, the winds properties depend on the SFR of the host galaxy, 
calibrated at low-$z$. Although this is a rather simplistic parametrization of 
a very complicated process, it introduces self-regulation in the galaxy
growth. This growth is based on the local, instantaneous SFR rather than being  
independent of host galaxy properties, as in the CW model.

In our simulations, the VW model exhibits the most complicated (e.g., nonmonotonic) 
behavior in terms of interplay between galactic 
outflows, gas consumption and SF among all the wind models. This behavior is especially
pronounced for the gas fraction, $f_{\rm gas,gal}$. On the average, 
the gas fraction in VW galaxies lies between the gas-poor NW galaxies and the CW gas-rich ones,
at the end of the simulations. This is also true for the median SFRs. The VW objects also show a 
strong decline of $f_{\rm gas,gal}$  
with the galaxy mass, $M_*$, in a sharp disagreement with the CW models.

When comparing with observational data (e.g., Figure\,\ref{fig:ssfr}), a case in favor of 
the VW prescription can be made. 
Due to the direct coupling between galaxy and wind properties imposed 
in the VW prescription, the SFR is correlated with the mass loss rate.
This self-regulation is very important for the intermediate
mass galaxies, $10^{8.5} < M_* < 10^{10}$, which are the most common
population at these high $z$, and probably the main contributors to
reionization, but see \citet[][]{Madau.Haardt:15}. We find that VW galaxies exhibit
a tight correlation sSFR -- $M_*$ which lies within $1\sigma$ from the observationally determined
median shown in Figure\,\ref{fig:ssfr}.

Observational results concerning the state of the IGM metallicity 
at high-$z$  make it difficult to draw conclusions 
in favor of specific wind models in identical environment, based on the differences between 
their efficiency in spreading metals on large-scales. 
Recent studies focusing 
on the statistics of C\,IV absorbers in spectra of distant quasars have shown 
a decline in their occurrences for $z > 5.3$ \citep{Becker.etal:09,Ryan-Weber.etal:09}. 
\citet{Ryan-Weber.etal:09} found a cosmic mass density of C\,IV ions at $z\sim 5.7$ of 
$\Omega_{\rm CIV} = 4.4 \times 10^{-9}$, a factor of 3.5 lower than the corresponding 
value for lower redshifts, and derived a corresponding lower limit for 
the IGM metallicity of $Z_{\rm IGM} \ga 10^{-4}\,Z_{\sun}$. 

The enrichment levels 
in the IGM at $z\sim 6$ in all of our simulation runs seems to satisfy this criterion 
based on the IGM metallicity maps shown in Fig.\,\ref{fig:tmaps}. Even the 
NW model is able to reach peak metallicities of $Z\sim 10^{-3}\,Z_{\sun}$ 
in the IGM (Fig.\,\ref{fig:metal_distrib}) despite the clear lack of metal 
spread on large-scales and a very small volume filling factor (Fig.\,\ref{fig:tmaps}). 
The reason for the relatively efficient 
IGM pollution even in the NW is likely caused by the accelerated evolution due to the 
presence of the overdensity which boosts SFR and, consequently, the metal production 
in this region. 

It is worth noting that other simulations used different outflow prescriptions to model
the observed IGM metallicity at low $z$ \citep[e.g.][]{Oppenheimer.etal:11}.
These of course do not include the no-wind model.
\citeauthor{Oppenheimer.etal:11} found that C\,IV absorbers are associated with T$\sim 10^4\,K$ 
gas in halos for which, in our models, this material is found at similar 
metallicity levels (Fig.\,\ref{fig:Z-T-rho_maps}).

The metal content of the IGM is affected as well by the presence of the 
overdensity because of the development of massive galaxies. 
This is in contrast with the average density (UCRs) models
where no metal-rich outflows are observed. Nevertheless, we do find that the peak 
metallicity attained in few regions in the IGM of the UCW run (visible as high-metallicity 
spots in Fig. \ref{fig:TZmaps_UCR}) 
is on a similar level as the CW model. So, even though the galaxy 
population is less developed, the metallicity content is of the same order, 
albeit in limited volumes --- a direct consequence of galaxy evolution. When averaged
over the computational volume, the IGM metallicity in UCR models is lower than the CR
models, with exception of NW.

Heating the IGM to $T\ga 10^5$\,K will destroy the dust
and facilitate reionization at much earlier times, being in
contradiction with the latest reionization time estimates \citep{Becker.etal:15,McGreer.etal:15}. 
Furthermore, the H\,I opacity would decrease rapidly from even
earlier redshifts, in contradiction with the results from \citet{Fan.etal:06}. 
At the same time, the IGM metallicity is observed to be very high in our CW models,
nearly sub-solar, which should leave an imprint in the spectra of the
high-$z$ objects, such as quasars and Lyman\,$\alpha$ emitting galaxies. 
However, this has not been detected yet. Furthermore, it will also help the SF process at
later stages, favoring the formation of proto-galaxies in the
surroundings in large numbers.  Therefore, galaxies, such as the so-called CR7 galaxy 
\citep[e.g.][]{Sobral.etal:15}, could be found in vicinity of similar objects, forming 
proto-cluster regions \citep[e.g.,][]{Trenti.etal:12}. 

Although one could state that these are mere results coming from a high-density
environment, this situation is not very different when comparing the
metallicity levels between average and high-density
environments. Their respective overall metallicity profiles lie within
0.5\,dex, but their distributions for the different media
(i.e., the IGM, halo gas and the ISM) are very similar (Fig.\,\ref{fig:metal_distrib}). 
Their overall temperature profiles are similar as well (Fig.\,\ref{fig:TZ_profiles}).

Overall, the temperature in the VW model lies within $10^4$\,K
and metallicity at log\,$(Z/Z_{\odot})\approx -2$, a much more acceptable
range which is also in agreement with reionization.
Figure\,\ref{fig:ssfr} shows that 
the VW model is able to reproduce both the observed 
sSFR evolution between $z\sim 8$ and $z\sim 6$ and the sSFR-stellar 
mass relation at $z\sim 6$. The CW and NW models fail to satisfy simultaneously 
these two constraints. Moreover, the NW run shows a clear disagreement with observed levels of 
star formation at $z\sim 8-6$. 
At $z\sim 6$, the scaling of sSFR with stellar mass is underestimated over the 
entire mass range $10^{8.8}-10^{10}\,\hmsun$ spanned by \citet{Salmon.etal:15} observations 
(Fig.\,\ref{fig:ssfr}, left). The sSFR values are also off at $z\sim 8$ and $z\sim 7$, not 
only for the entire galaxy population, but even when 
considering galaxies only within the observed stellar mass range. 

Unfortunately, computational expenses prohibit implementing high-resolution studies 
in large-scale cosmological simulations.
The full cosmological simulation still needs to rely on a rather simple scaling.
As we discussed in section\,\ref{sec:winds}, the VW model possesses one important feedback 
parameter --- 
the wind velocity dependence on the host galaxy SFR, which constrains the wind properties 
better than 
the CW model. Although its value is tuned-up to the low-$z$ observation \citep{Martin:05},
the median VW speed appears lower than that of the CW.
As a result, the VW does not overheat the IGM compared with the CW, while it
regulates the SF and galaxy growth, and shows an agreement with the sSFR observations. 
It implies that the scaling relation that the VW adopts still holds in galaxy evolution at 
$z\ga 6$ and the predictions 
from the VW simulation, such as the galaxy mass function, SFR, sSFR, and $f_{\rm gas,gal}$ 
would be more reliable than those in CW. 

The low SF levels at $z < 7$ in the NW case are caused by the absence of winds which 
allows efficient consumption of gas by SF at high-$z$ and 
gradually depletes galaxies of their gas. This also causes metallicity in galaxies 
to reach very high values by $z\sim 6$ for a broad range of stellar masses 
(Fig.\,\ref{fig:halogas_Mhalo_Z}). In particular, for low-mass galaxies, we find 
unrealistic metallicity levels in the stellar and gaseous components, causing 
the resulting mass-metallicity relation to be nearly flat in the NW case, as 
opposed to increasing with mass as shown by the other two wind models. 
Furthermore, the CW model is able 
to reproduce the observed sSFR but fails at capturing correctly 
the decreasing trend of sSFR with stellar mass, as it 
shows a positive correlation between those two quantities. 
in addition, the CW model appears in contradiction with observations 
at lower redshifts, such as an excessive metal enrichment of low density gas 
\citep[e.g.,][]{Oppenheimer.Dave:08,Choi.Nagamine:11}.
We, therefore, conclude that the VW wind model more closely reproduces galactic wind,
IGM- and halo-gas properties at high $z$, although a thorough analysis using radiative 
transfer is needed to confirm this conclusion.
 
In summary, we find that galaxy evolution at $\sim 6-12$ depends strongly on the type of
feedback mechanisms applied. While specific prescriptions for galactic winds allow us to
reproduce some of the observed properties of galaxies at these redshifts, the best fit is 
attained by using the VW model, which couples wind properties to the ongoing star formation 
and to the depth of the galactic potential well. No perfect wind model can be decided
at this moment, perhaps a new wind model is necessary which is more tightly related to the
underlying galaxy properties.
The VW model, although quite simplistic, 
appears best equipped to reproduce the galaxy properties, both in overdense and 
average regions in the universe. Future improvements of galactic wind models must include
dependency on morphological and radiative properties of host galaxies.

\acknowledgments

We thank Volker Springel for providing us with the original version of GADGET-3.
We are grateful to Ken Nagamine for fruitful discussions, to Steven Finkelstein, 
Richard Bouwens 
and Hidenobu Yajima for helpful comments, and to Yehuda Hoffman for producing the initial 
conditions by means of Constrained Realizations. 
This work has been partially supported by the NSF grant AST-080776, the HST/STScI grant 
AR-12639.01-A and by JSPS KAKENHI grant \#16H02163 (to I.S.). R.S. is partially 
supported by NSF grant AST-1208891. 
I.S. is grateful for support from International Joint Research 
Promotion Program at Osaka University.
J.H.C acknowledges support from NASA ATP NNX11AE09G, 
NSF AST-1009799, and Caltech/JPL SURP Project No. 1515294 through the UT Austin 
(P.I. Paul Shapiro). E.R.D. thanks DFG for support under SFB 956. Support for
HST/STScI grant was provided by NASA through a grant from the STScI, which is 
operated by the AURA, Inc., under NASA contract NAS5-26555. Simulations have been
performed on the University of Kentucky DLX Cluster and using a generous allocation 
on the XSEDE machines to I.S.



\end{document}